\documentclass[twocolumn, floatfix]{aastex631}
\usepackage{ulem}
\usepackage{float}
\usepackage{gensymb}
\usepackage{array}
\usepackage{booktabs} % Para as linhas horizontais

\usepackage{breqn}
\usepackage[flushmargin]{footmisc}

\usepackage{amsmath}

\shorttitle{}
\shortauthors{Vicentin et al.}

\graphicspath{{./}{figures/}}

\begin{document}

\title{Selecting Clusters and Protoclusters via Stellar Mass Density: II. Application to HSC-SSP Observations
}

\correspondingauthor{Vicentin, Marcelo C.}
\email{marcelo.vicentin@usp.br}

\author[0000-0002-9191-5972]{Marcelo C. Vicentin}
\affil{Universidade de S\~ao Paulo, Instituto de Astronomia, Geof\'isica e Ci\^encias Atmosf\'ericas, Departamento de Astronomia, \\ SP 05508-090, S\~ao Paulo, Brasil}
\affil{Department of Astrophysical Sciences, Princeton University, Peyton Hall, Princeton, NJ 08544, USA}

\author[0000-0002-3876-268X]{Laerte Sodré Jr.}
\affil{Universidade de S\~ao Paulo, Instituto de Astronomia, Geof\'isica e Ci\^encias Atmosf\'ericas, Departamento de Astronomia, \\ SP 05508-090, S\~ao Paulo, Brasil}

\author[0000-0002-0106-7755]{Michael A. Strauss}
\affil{Department of Astrophysical Sciences, Princeton University, Peyton Hall, Princeton, NJ 08544, USA}

\author[0000-0002-6268-8600]{Erik V. R. de Lima}
\affil{Universidade de S\~ao Paulo, Instituto de Astronomia, Geof\'isica e Ci\^encias Atmosf\'ericas, Departamento de Astronomia, \\ SP 05508-090, S\~ao Paulo, Brasil}

\author[0000-0003-2860-5717]{Pablo Araya-Araya}
\affil{Universidade de S\~ao Paulo, Instituto de Astronomia, Geof\'isica e Ci\^encias Atmosf\'ericas, Departamento de Astronomia, \\ SP 05508-090, S\~ao Paulo, Brasil}

\begin{abstract}

We present a selection of candidates of clusters and protoclusters of galaxies identified in the photometric data of the HSC-SSP Wide Public Data Release 3 (PDR3), spanning the redshift range $\rm 0.1 \leq z \leq 2$. The selection method, detailed in \citet{vicentinI25}, involves detecting massive galaxies located in high-density regions of matter, identified as potential central dominant galaxies, i.e., (proto)BCGs. Probabilistic criteria based on proximity to the candidate central galaxy and the expected stellar mass of member galaxies are applied to identify likely members of each structure. We produced updated photometric redshift estimates using deep learning methods trained on a dataset combining spectroscopic redshifts from the HSC-SSP Wide PDR3, high-accuracy photometric redshifts from the COSMOS2020 catalog, and mid-infrared data from the unWISE catalog for matched sources. Our method achieves a predicted purity of $\sim 90\%$ in detecting (proto)clusters, with $\gtrsim 65\%$ correctly identifying the (proto)BCG. A total of 16,007 candidate (proto)clusters were identified over an effective area of $\rm \sim 850 \ deg^{2}$ within the HSC-SSP Wide footprint. Comparisons with other existing catalogs reveal a good level of consistency, while also highlighting that different methods yield complementary discoveries. We further compare richness and halo masses from our optical catalog with those from recent X-ray cluster catalogs (eROSITA and MCXC-II), finding a moderate positive correlation and a scatter of $\rm \sim 0.4$ dex. This catalog provides a valuable new set of targets for the Prime Focus Spectrograph (PFS) instrument.

\end{abstract}

%-------------------------------------------------------------------
\section{Introduction} \label{sec:intro}

The gravitational collapse of the earliest perturbations in the primordial density field created the seeds that eventually developed into the largest gravitationally bound structures in today’s universe: galaxy clusters. Following this collapse, according to the hierarchical structure formation scenario \citep{white78}, smaller objects formed at the centers of potential wells, leading to the creation of the first stars and, subsequently, the first galaxies. Even during the early stages of the Universe, at $\rm z \gtrsim 5$, groups of galaxies are already observed to be sufficiently close to one another that they will merge into a single virialized structure with a total mass of $\rm M \gtrsim 10^{14} \ M_{\odot}$ in the future \citep[e.g., ][]{overzier09, lovell18, toshikawa24, stawinski24}. These structures are the progenitors of galaxy clusters, known as protoclusters \citep{overzier16}. Building a deep and complete sample of galaxy clusters and protoclusters across large areas of the sky is important as it provides a pathway to constrain cosmological parameters \citep[e.g., ][]{abdullah23, chiu24, sunayama24} and to understand the evolution of galaxies in dense environments \citep[e.g., ][]{ebeling14,castignani22,hartzenberg23,taamoli24}.

In recent decades, wide-area optical multi-band photometric surveys \citep[e.g.,][]{york00, chambers16, mendesdeoliveira19, bonoli21} have become essential tools for identifying galaxy cluster and protocluster candidates, providing extensive datasets that enable systematic searches for large-scale structures across vast cosmic volumes. By leveraging photometric measurements, it is possible to estimate quantities such as photometric redshifts; thus, searching for galaxy overdensities within specific redshift slices can reveal the presence of virialized clusters or forming protoclusters. While numerous studies have focused on clusters at low to intermediate redshifts \citep[e.g.,][]{takey11, wh12, bellagamba18, aguena21, werner23, doubrawa24}, and others have targeted high-$z$ protoclusters ($z > 2$) in deep but narrow surveys \citep[e.g.,][]{higuchi19, toshikawa20, hu21}, the redshift range $1 < z < 2$ remains relatively underexplored. Some efforts have aimed to fill this gap, most notably the Massive and Distant Clusters of WISE Survey \citep[MADCOWS; e.g.,][]{stanford14, gonzalez19, thongkham24, thongkham24_b}, which combines infrared and optical data to identify massive clusters in this transitional epoch. Nonetheless, large-area, homogeneous cluster catalogs in this redshift range are still limited, motivating the development of complementary approaches.

In this study, we aim to identify galaxy (proto)cluster candidates in the wide layer of the Hyper-Suprime Cam Subaru Strategic Program Public Data Release 3 \citep[HSC-SSP PDR3;][]{aihara22} within the redshift range $0.1 < z < 2$. We estimate new photometric redshifts by incorporating spectroscopic redshifts available for objects within the HSC-SSP PDR3  footprint, adding high-accuracy photometric redshifts from the COSMOS2020 catalog \citep{weaver22}, and integrating mid-infrared band information from W1 and W2 in the unWISE catalog \citep{schlafly18} for objects with matching counterparts. To select candidates, we follow the methodology outlined in \citep{vicentinI25}, which begins by identifying the dominant galaxy within each structure. A brief review of this algorithm is presented in Section \ref{sec: review}.

Dominant/brightest (proto)cluster galaxies, or (proto)BCGs, are massive and often found at the peak of the local galaxy density field, suggesting a strong correlation between local density and the probability of a given massive galaxy being a BCG. As shown in V25a, there is a higher incidence of structures in the protocluster stage at redshifts $\rm z > 1.5$, representing the progenitors of present-day galaxy clusters that have not yet reached their final mass assembly but will evolve into massive clusters ($\rm \log(M_{halo}/M_{\odot}) \gtrsim 14$) at lower redshifts \citep[for a review, see][]{overzier16}. However, by applying our methods to identify (proto)BCGs in a set of mocks that emulate HSC-SSP Wide observations \citep{araya24}, we demonstrate that it is possible to select a sample of (proto)BCG candidates with a purity of $\gtrsim 65\%$ even in this redshift range.

Other algorithms of this nature have already been applied to the HSC-SSP Wide survey area. The CAMIRA algorithm \citep{oguri14} was developed based on the strategy of identifying concentrations of galaxies that are potential members of the red sequence at a given redshift. Using this approach, CAMIRA identified 20,654 galaxy cluster candidates in the HSC-SSP Wide Public Data Release 3 (PDR3) within the redshift range $\rm 0.1 < z < 1.4$. \citet{wh21} performed a cross-match between the HSC-SSP Wide PDR2 and the unWISE catalog to incorporate mid-infrared measurements in the WISE W1 (3.6 $\mu$m) and W2 (4.5 $\mu$m) bands, considering only objects with counterparts in this catalog. Using the galaxies' photometry to estimate photometric redshifts, they identified 21,661 cluster candidates within $0.1 < z < 2$, selecting concentrations of galaxies located within the same redshift slice.

It is important to highlight that different selection criteria and techniques for identifying (proto)cluster of galaxies candidates lead to different outcomes, with the catalogs produced by these algorithms complementing one another. In the present work, we emphasize several distinguishing features of our approach compared to others. Notably, our selection does not explicitly assume any galaxy color cut, nor do we limit our sample to objects with counterparts in the unWISE catalog. Furthermore, we extend our candidate selection up to redshift 2, leveraging newly photometric redshift estimates. Unlike \citet{wh21}, who only estimated photo-z for galaxies with unWISE counterparts, our estimates were computed for galaxies independently of infrared detection, using a distinct methodology. The construction of a new catalog of (proto)cluster candidates up to high redshifts is further justified as it provides a valuable target list for observations with the \textit{Subaru Prime Focus Spectrograph} \citep[PFS; for an overview, see][]{tamura16}.

This work is structured as follows: In Section \ref{sec: data}, we provide a detailed description of the HSC-SSP Wide PDR3 data used in this study. Section \ref{sec: photozs} describes the procedures adopted to estimate photometric redshifts. Section \ref{sec: cluster_sel} presents the catalog of (proto)cluster candidates selected in this study, as well as comparisons with catalogs produced by CAMIRA \citep{oguri18}, \citet{wh21} (hereafter WH21), and recent catalogs of galaxy clusters identified through extended X-ray emission \citep{sadibekova24, bulbul24}. Lastly, we summarize our findings and conclusions in Section \ref{sec: conclusions}. Throughout this work, we adopt a $\rm \Lambda$CDM concordance cosmology with $\rm h = 0.673$, $\rm \Omega_{\rm m} = 0.315$, and $\rm \Omega_{\Lambda} = 0.685$ \citep{planck1}.

%-------------------------------------------------------------------
\section{Data} \label{sec: data}

Our goal is to select galaxy cluster and protocluster candidates using the algorithm described in V25a, in the field covered by the HSC-SSP Wide PDR3 photometric survey \citep{aihara22}. In this section, we present a brief description of this dataset. %We included details about the spectroscopic data incorporated in the HSC-SSP database, as well as information from other auxiliary catalogs that we will utilize in Section \ref{sec: photozs}.

The HSC-SSP \citep{aihara18} is an optical imaging survey designed to investigate a wide range of astrophysical questions, spanning from cosmology to the dynamics of solar system bodies. The survey employs the Hyper Suprime-Cam \citep[HSC;][]{miyazaki18}, a wide-field imaging camera mounted at the prime focus of the Subaru 8.2m telescope. The Wide layer of the survey, combining all three data releases, covers approximately 1200 deg² of the sky, offering complete photometry in five broad-band optical filters \citep[grizy;][]{kawanomoto18}, with a depth reaching up to $i \sim 26$. % 670 deg^2 are new imaging from pdr3; 1200 deg^2 with complete photometry in grizy, otherwise ~1470deg^2

We utilized \texttt{cmodel} magnitudes along with their corresponding errors, both processed using hscPipe v8, a customized version of the LSST Science Pipelines \citep{juric17, bosch18, bosch19, ivezic19}. To ensure the reliability of the photometric measurements—avoiding artifacts from bad pixels and low signal-to-noise detections—we applied a magnitude limit of $i < 25$, corrected for extinction due to Milky Way dust. Additionally, we excluded galaxies with any of the following flags in any of the measured bands (similar to \citealp{tanaka18}):
\begin{enumerate}
    \setlength\itemsep{0.1em}
    \item \texttt{grizy_cmodel_flag}
    \item \texttt{grizy_pixelflags_edge}
    \item \texttt{grizy_pixelflags_interpolatedcenter}
    \item \texttt{grizy_pixelflags_saturatedcenter}
    \item \texttt{grizy_pixelflags_crcenter}
    \item \texttt{grizy_pixelflags_bad}
    \item \texttt{grizy_sdsscentroid_flag}
\end{enumerate}

To eliminate duplicates, we applied the criterion \texttt{isprimary = True}. Additionally, we selected only those objects with \texttt{grizy_extendedness_value} = 1, which are classified as extended sources in the HSC-SSP database.

We also applied photometric corrections provided in the PDR3. One of these corrections concerns the transmission of the $i$ and $r$ filters. As described in \citet{aihara22}, the $i$-band filter initially used in the survey exhibits significant radial variation in its transmission curve, resulting in a donut-like pattern in the background of images. To resolve this, a new filter, $i2$, was introduced. Consequently, some survey images were taken with the original $i$ filter, others with $i2$, and some with both. Although the $i$ and $i2$ filters have similar transmission curves, they are not identical, leading to variations in the photometry of objects across these regions. A similar issue occurs with the $r$ filter. To address these inconsistencies, we used the corrections available in the HSC-SSP Wide database to convert the $r$ and $i$ measurements to $r2$ and $i2$, respectively.

%Another correction made in PDR3 concerns the photometric calibration, which was performed using the Forward Global Calibration Method \citep[FGCM;][]{burke18}. There is a slight inhomogeneity in this calibration, partly due os novos filtros $r2$ e $i2$, discutidos no parágrafo anterior. To address this, after applying these corrections, color-color diagrams of the observed stars were used to estimate average offsets from the expected positions for each filter. Accordingly, we also applied these offsets to objects magnitudes.

Finally, we also included a bright star flag \citep[\texttt{grizy_mask_brightstar_any, }][]{coupon18}. Candidate dominant galaxies within a candidate (proto)cluster, i.e., BCGs or protoBCGs, are required to not have this flag set (see Section \ref{sec: catalog}).

%-------------------------------------------------------------------
\section{Photometric redshifts}
\label{sec: photozs}

Photometric redshift estimation is a crucial technique in large photometric surveys. By estimating redshifts from multi-band photometry, it is possible to efficiently study vast numbers of galaxies across the universe, allowing, for example, the exploration of cosmic evolution and the identification of galaxy clusters, the goal of the present work. This method significantly expands the reach of surveys, making it possible to conduct wide-field studies of the universe’s structure and history, specially in regions of the sky where spectroscopic data is sparse or unavailable. 

%We used the photometric redshifts from COSMOS2020 obtained through the template fitting code \texttt{LePhare} \citep{ilbert06}, using additional measurements in more than 40 filters from UV to mid-IR. The dispersion ($\sigma_{NMAD}$) and the outlier fraction (both metrics defined as we are defining in Section \ref{sec: photoz_result}) for galaxies with $\rm \textit{i} < 22.5$ are both below 1\%. Even for the faintest magnitude bin, $25 < \textit{i} < 27$, the accuracy reaches 4\% and the outlier fraction is approximately 20\%.

The HSC-SSP provides value-added catalogs with photometric redshift estimates using various techniques \citep{tanaka18, nishizawa20}, including those based on template-fitting and machine learning. However, for redshifts higher than $\rm z \sim 1.4$, key spectral features like the 4000 {\AA} break move out of the photometric coverage of the survey's \textit{wide} layer (\textit{grizy} bands). This loss of information leads to a decline in the quality of the estimates derived from this photometric system. Additionally, the number of objects with spectroscopic redshifts decreases significantly, which limits the representativeness of spectral models for template fitting and the construction of training samples for machine learning applications.

In this work, we will estimate photometric redshifts using deep learning techniques similar to those described in \citet{lima22}. To build a training sample that aligns with our objectives (selecting (proto)cluster candidates up to $\rm z \sim 2$), we will adopt two strategies:

\begin{itemize}

    \item Incorporate galaxies with high-accuracy photometric redshifts from the COSMOS2020 catalog \citep{weaver22}, particularly to increase the number of objects with redshift information at higher redshifts;

    \item Include measurements in the W1 (3.6 $\mu$m) and W2 (4.5 $\mu$m) bands from the unWISE catalog \citep{schlafly18} for cross-matched galaxies.

\end{itemize}

In the following subsections, we will provide details on how the spectroscopic sample of the HSC-SSP was constructed; how we incorporated information from other catalogs into the original spectroscopic sample; the technique we adopted to de-bias the final training sample; the deep learning network architecture applied for the estimates; and finally, the results we obtained by comparing our estimates with the spectroscopic redshifts from a blind test sample.

\subsection{Public spectroscopic redshift in HSC-SSP database}

The spectroscopic sample used in this work is a collection of observations compiled from the literature and made available in the HSC database\footnote{https://hsc-release.mtk.nao.ac.jp/}. We used the most recent sample available, which was provided in the Public Data Release 3 (PDR3). The spectroscopic samples are available for different layers, along with their respective photometry. Since our focus is on selecting galaxy cluster candidates in the wide layer, we used the sample containing galaxies within this footprint and photometry matching that depth. This dataset was constructed from the following samples: zCOSMOS DR3 \citep{lilly09}, 3D-HST \citep{skelton14, momcheva16}, SDSS DR15 \citep{aguado19}, GAMA DR3 \citep{baldry18}, UDSz \citep{bradshaw13, mclure13}, VANDELS DR2 \citep{pentericci18}, C3R2 DR2 \citep{masters2}, VVDS \citep{lefevre13}, DEIMOS 2018 \citep{hasinger18}, FMOS DR2 \citep{silverman15}, LEGA-C DR2 \citep{straatman19}, PRIMUS DR1 \citep{coil11, cool13}, VIPERS DR2 \citep{garilli14}, WiggleZ DR1 \citep{drinkwater10}, and DEEP23 DR4 \citep{davis03, cooper11, cooper12, newman13}. \citet{tanaka18} and \citet{nishizawa20} show which flags were applied in the public spectroscopic redshift catalogs. 

The photometric corrections described in Section \ref{sec: data} were also applied to these data. After this selection, this spectroscopic catalog has a total of 741288 galaxies.

\subsection{COSMOS2020 and unWISE information}

As previously mentioned, our strategy to improve photometric redshift estimates at high redshifts, particularly for $\rm z > 1.4$, involves incorporating galaxies with highly accurate photometric redshifts from the COSMOS2020 catalog into the spectroscopic sample and adding W1 and W2 band magnitudes for galaxies listed in the unWISE catalog.

For the COSMOS2020 objects, we identified which galaxies with estimated redshifts in this catalog were not included in the spectroscopic sample. For these galaxies, we performed a cross-match with the HSC-SSP wide layer database to include its photometry, since COSMOS2020 provides photometry from the deep layer. This step ensures that these galaxies have consistent photometry with the spectroscopic sample. 

After adding 80,000 COSMOS2020 galaxies, we conducted a cross-match with the unWISE catalog using a maximum distance of 2.75 arcseconds, as recommended by \citet{schlafly18}. Out of the 820,634 galaxies in the combined spectroscopic and COSMOS2020 catalog, W1 band magnitudes were added for 380,134 galaxies (46\%), and both W1 and W2 band magnitudes were added for 287,846 galaxies (35\%). 

From this sample, we randomly selected 100,000 galaxies, splitting them evenly into a test set and a validation set. This approach ensures a blind sample for applying the neural network and assessing the accuracy of the photometric redshift estimates (Section \ref{sec: photoz_result}).

\subsection{Unbiasing the training sample}
\label{sec: unbias}

Another necessary step in constructing the final training sample is the unbiasing of the spectroscopic sample. The spectroscopic sample is assembled from a collection of observations with different scientific goals. As a result, this is a heterogeneous sample, which could introduce biases during the machine learning training process. In supervised models, the training data selection injects prior knowledge into the learning process by design. Assume you trained a model $\mathcal{M}$ (which has parameters $\theta$) for classification or regression, using a training set $\mathcal{D}$. The distribution of a new observation $y$ is

\begin{equation}
    P(y|\mathcal{D}) = \int P(y, \theta|\mathcal{D}) d\theta = \int P(y|\theta, \mathcal{D}) P(\theta) d\theta,
\end{equation}

showing that a prediction indeed depends implicitly on the model and training set.

Additionally, often the test set is drawn from the same data sample as the training set. In this case, good results may indicate that the model is directly fitting the bias in the training set (confirmation bias). This may lead to bad results when the data set on which the models will be applied have properties distinct from those of the training set. This is the case for photometric redshifts, when the training sets comprise many different sources, covering different redshifts and with different selection criteria, and the algorithm is applied to, say, a magnitude-limited sample. This is called sample selection bias \citep[e.g.,][]{kremer17}.

To address this problem, we adopted the following method: Suppose we have a true sample ($Tr$), where the real redshift distributions as a function of magnitude are known. We can obtain approximately unbiased distributions from a biased training sample ($BTS$) by randomly sampling the $Tr$ and selecting in $BTS$ the galaxy with closest spectroscopic redshift and magnitudes to the sampled galaxy. In other words, the galaxy selected from $BTS$ for the unbiased training set ($UTS$) is almost a clone (i.e., similar redshift and photometric properties) of the galaxy sampled from $Tr$.

In practice, we use the COSMOS2020 catalog--which has similar depth to the HSC-SSP Wide observations--as the true redshift distribution, denoted as $Tr$. From this catalog, we performed sampling in the magnitude range $18 \leq i \leq 25$, divided into bins of $\Delta i = 0.2$. Within each bin, for every galaxy in the $BTS$, we selected an equal number of galaxies that are the nearest neighbors in the five-dimensional HSC-SSP Wide magnitude space ($grizy$) to those in $Tr$, constrained to lie within the same redshift interval $\Delta z_{i} \equiv [z_{i} \pm 0.05]$, where $z_{i}$ is the redshift of the $i$-th target galaxy being matched. The redshift bin was chosen within the COSMOS2020 photometric redshift accuracy to provide reliable clones and, at the same time, to assure the proper redshift sampling. Notice that there will be repetitions in the selection of these galaxies, so the number of repetitions in the final unbiased sample, i.e., $UTS$, will act as a weight for each object. %Figures \ref{fig: uts_cosmos_1} and \ref{fig: uts_cosmos_2} show the distribution of sampled galaxies and COSMOS2020 for different $\Delta i$.

\subsection{Undersampling low-z galaxies}
\label{sec: unders}

Another procedure that resulted in a significant improvement in photo-z estimates at high redshifts, particularly for $\rm z > 1.4$, without compromising the results for photo-$z$'s below this value, was to reduce the number of galaxies in the training sample within the redshift range $\rm 0 < z < 1.4$. 

\begin{figure}

	\includegraphics[width=\columnwidth]{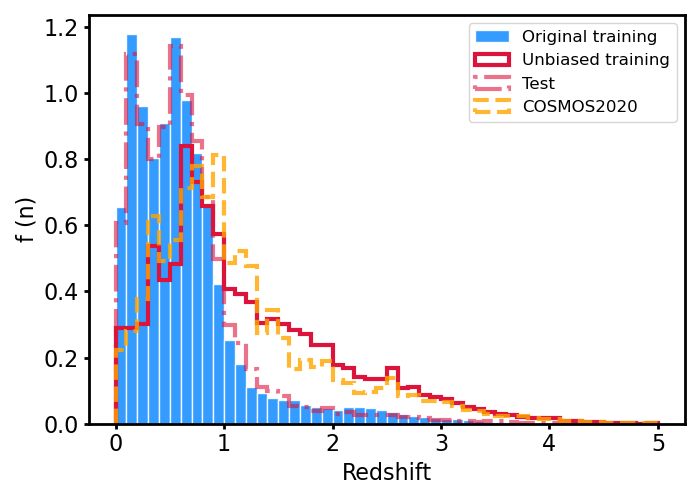}
    \caption{The blue histogram represents the redshift distribution of the original training sample, consisting of spectroscopic data and photometric redshifts from COSMOS2020. The dot-dashed histogram indicates the distribution of the test sample. The red histogram shows the distribution of the unbiased training sample (Sections \ref{sec: unbias} and \ref{sec: unders}). The yellow dashed histogram represents the photometric redshift distribution from COSMOS2020. All histograms are normalized so that the total area sums to unity.
}
    \label{fig: training_cosmos_dist}
\end{figure}

To achieve this, we used the number of galaxies within $\rm 1.45 < z < 1.5$ as a reference number ($\rm n_{ref}$). For bins (of width $\Delta z =$ 0.05) below this redshift that contain more objects than $\rm n_{ref}$, we randomly removed 90\% of the objects exceeding $\rm n_{ref}$. Consequently, for these bins, the number of objects becomes $\rm n_{ref} + 0.1 \times n$, where $\rm n$ is the total number of galaxies in a given redshift bin. After applying this selection, the total number of galaxies in the training sample is approximately 260,000. The test sample, with 50,000 galaxies, therefore represents about 20\% of this total. The distribution of the final training sample (solid red line), along with the original spectroscopic sample (blue bars), the test sample (dot-dashed light red line), and the photometric redshifts from COSMOS2020 (dashed orange line), are presented in Figure \ref{fig: training_cosmos_dist}.

In the next subsection, we will present the neural network method adopted to obtain a model for predicting photometric redshifts, along with an evaluation of the results.

\subsection{Bayesian neural network and photo-z results}
\label{sec: photoz_result}

Multilayer perceptron (MLP) neural networks \citep[e.g., ][]{haykin98} consist of an input layer, multiple hidden layers, and an output layer, which, in this study, predicts photometric redshifts. Each neuron in a layer is connected to every neuron in adjacent layers, forming a fully connected feed-forward network. The hidden layers iteratively adjust connection weights using backpropagation to minimize a loss function, improving prediction accuracy by aligning outputs with true values.

%To prevent overfitting, model performance is monitored using a validation set. If validation loss increases while training loss decreases, it indicates poor generalization. To address this, techniques like early stopping, regularization, and dropout are employed, ensuring the model effectively applies learned patterns to new data rather than memorizing the training set.

Bayesian neural networks \citep[BNNs - e.g.,][]{bishop97} extend this framework by treating network weights as probability distributions instead of fixed values. Unlike deterministic models that provide single-point predictions, BNNs incorporate uncertainty by using Bayes’ theorem to estimate posterior weight distributions. This probabilistic approach enables more robust predictions, accounting for uncertainties in both model parameters and data.

Here, we use a Bayesian Neural Network (BNN) where the network architecture parameters, such as the number of hidden layers, activation functions, and the number of neurons per layer, were optimized using the \texttt{Optuna} package \citep{akiba19}. The optimization aimed to find the configuration that minimized the variance in redshift estimates. The input data consists of galaxy magnitude measurements in the $grizyW1W2$ bands, along with the associated measurement errors. Consequently, the input layer is composed of 14 neurons. For galaxies lacking information in the $W1W2$ bands and their corresponding errors, we assign a value of 99. Neural networks can naturally interpret such out-of-range values as missing data due to the non-linearity of activation functions and weight optimizations during training, ensuring that the absence of information does not introduce biases into the model \citep{chollet18}.

\begin{figure*}

	\includegraphics[width=\textwidth]{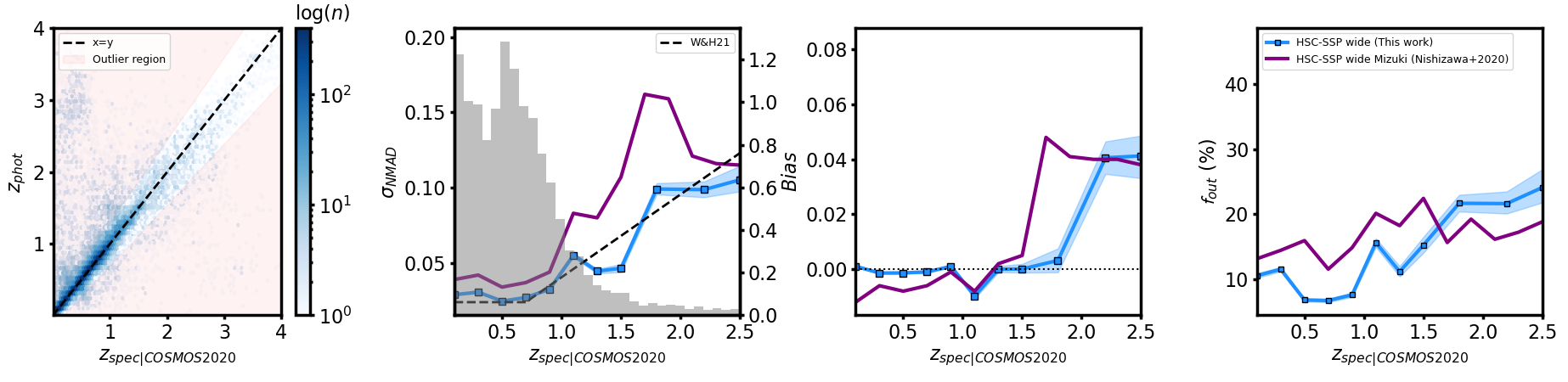}
    \caption{From left to right: Photometric redshifts ($\rm z_{phot}$), uncertainties $\sigma_{NMAD}$ (\ref{eq: nmad}), $Bias$ (\ref{eq: bias}), and outlier fraction $f_{out} ($\ref{eq: f_out}), respectively, as a function of spectroscopic or COSMOS2020 redshifts ($\rm z_{spec|COSMOS2020}$). Blue lines show the results obtained in this work, while purple lines the results obtained by \citet{nishizawa20}. The dashed line in the second plot denotes the result obtained by \citet{wh21}. 
}
    \label{fig: pz_test_samp}
\end{figure*}

Figure \ref{fig: pz_test_samp} illustrates the results of the blind test sample estimates. The first panel shows the estimated redshifts versus the spectroscopic or COSMOS2020 redshifts. 

To evaluate the quality of the estimates, we use three metrics. The Normalized Median Absolute Deviation \citep[$\sigma_{NMAD}$, ][]{hoaglin83} to measure dispersion, is given by:

\begin{equation}
    \sigma_{NMAD} (z_{spec}) = 1.48 \times median \left(\frac{\Delta z - median(\Delta z)}{1 + z_{spec}}\right),
    \label{eq: nmad}
\end{equation}

\noindent
where $z_{spec}$ is the spectroscopic or COSMOS2020 redshift, $z_{phot}$ is the estimated photometric redshift, and $\Delta z = z_{phot} - z_{spec}$. The bias to measure systematic deviations, is given by:

\begin{equation}
    Bias (z_{spec}) = median \left(\frac{\Delta z}{1 + z_{true}}  \right),
    \label{eq: bias}
\end{equation}

and the outlier fraction is

\begin{equation}
    f_{out} (z_{true}) = \frac{N_{out}}{N_{tot}} 
    \label{eq: f_out},
\end{equation}
\\

\noindent
where $N_{out}$ is defined as the number of objects that satisfies the condition $|\Delta z| / (1 + z_{true}) \geq 0.15$, and $N_{tot}$ is the total number of objects. 

Figure \ref{fig: pz_test_samp} shows the results obtained. Blue curves with markers represent our results for these metrics, while the purple lines show the results obtained by \citet{nishizawa20} using the template fitting method called Mizuki. The dashed line in the $\sigma_{NMAD}$ versus redshift plot represents the relation obtained by WH21, where the estimates were made considering only galaxies with complete photometry in the $grizyW1$ bands, i.e., excluding galaxies without counterparts in the unWISE catalog. Our results exhibit a comparable scatter to WH21 and show improvements across all three metrics relative to \citet{nishizawa20}, except for the outlier fraction at $z > 1.6$, which is approximately 3\% higher.

%-------------------------------------------------------------------
\section{Selected galaxy cluster and protocluster candidates} \label{sec: cluster_sel}

In this section, we describe our selection of (proto)cluster candidates. In Section \ref{sec: review}, we summarize the procedures adopted to select dominant galaxies, i.e., BCGs or protoBCGs, and subsequently other member galaxies of the structures, as described in detail in V25a. In Section \ref{sec: catalog}, we discuss some characteristics of the selected (proto)cluster candidates in the HSC-SSP Wide PDR3 field; Comparisons with other cluster finder algorithms applied to HSC-SSP Wide, i.e., CAMIRA \citep{oguri18} applied to PDR3 at $\rm 0.1 < z < 1.4$, and WH21 applied to PDR2 at $0.1 < z < 2$ in Sections \ref{sec: cfs_comp}, \ref{sec: match_unmatch}, \ref{sec: highz_samp}, and \ref{sec: members_props}. Finally, in Section \ref{sec: xray}, we present comparisons between the optically selected clusters with counterpart in x-ray cluster catalogs.

\subsection{Review of the cluster finder}
\label{sec: review}

The galaxy cluster finder used in this work aims to identify first the dominant galaxies, which are typically located at the centers of galaxy structures where the galaxy density field peaks. It uses local stellar mass density measurements around pre-selected massive galaxies to find candidates for BCGs or protoBCGs. It calculates the stellar mass density within a cylindrical volume around each pre-selected candidate, considering galaxies located within a radius of 1 Mpc and a height determined by the redshift slice, which accounts for photometric redshift uncertainties. The volume is divided into three concentric rings, with equally spaced radial intervals, and a weighting factor based on the inverse radial distance is applied to prioritize galaxies closer to the candidate.

To measure the density contrast, the algorithm compares the calculated local density to a reference density averaged over a large area of the sky. This contrast is used to model the probability of a galaxy being a true dominant galaxy based on comparisons with mocks that emulate the HSC-SSP wide survey \citep{araya24}. The probabilistic models derived from simulations allows for the identification of dominant galaxy candidates in real observational data.

The algorithm identifies cluster member galaxies by evaluating the likelihood of their membership based on their distance to the dominant galaxy and the expected stellar mass, while accounting for contamination from background and foreground sources. Galaxies that have a higher probability of being genuine cluster members, as opposed to interlopers, are selected accordingly. The number of galaxies identified in this manner serves as a richness ($\lambda$) estimate for each cluster, which is then utilized to establish halo mass-richness relations. Unlike traditional cluster-finding methods, this approach focuses on locating dominant galaxies initially, without relying on any intrinsically color-based criteria, making it a more flexible tool for cluster identification, specially at higher redshifts, where protoclusters dominate, which tend to be composed of more star-forming blue galaxies than are mature structures.

\subsection{Galaxy (proto)cluster candidates catalog}
\label{sec: catalog}

We selected 16,007 galaxy (proto)cluster candidates in the redshift range $\rm 0.1 < z < 2$ across nearly the entire area of the HSC-SSP Wide Survey PDR3\footnote{For redshifts lower than $z \sim 0.1$, the typical photometric redshift uncertainties become comparable to the redshift itself, making it difficult to define precise redshift slices for structure identification.}. We use spectroscopic and COSMOS2020 photometric redshifts for those objects that have one of these measurements, prioritizing the spectroscopic data. Prior to applying our algorithm, we performed a visual inspection of the sky coverage and data quality across the survey, excluding regions with sparse photometric coverage or artifacts—primarily located near the survey edges. As a result, the effective detection area was reduced to approximately $\sim 850 \ \mathrm{deg}^{2}$. 

The number of selected candidates corresponds to the expected number of galaxy clusters for this area, divided into redshift bins within the total range of interest, as determined through mock analyses presented in V25a. The selected structures are those in which the dominant galaxies were chosen based on the highest probabilities. Importantly, while the selection is designed to match the expected number of galaxy clusters, it does not explicitly exclude protoclusters. As shown in V25a, at higher redshifts, a significant fraction of the selected structures are still in the protocluster stage. According to our definition, galaxy clusters are structures residing in halos with masses $\rm M_{halo} \geq 10^{14} \ M_{\odot}$, while a protocluster refers to a structure that has not yet reached this mass threshold but will do so at some future point. However, since both clusters and protoclusters represent genuine overdense regions in the large-scale structure, we do not classify the presence of protoclusters as contamination, but rather as a natural outcome of the selection process, reflecting how these structures are classified based on their evolutionary stage.

\begin{figure}

	\includegraphics[width=\columnwidth]{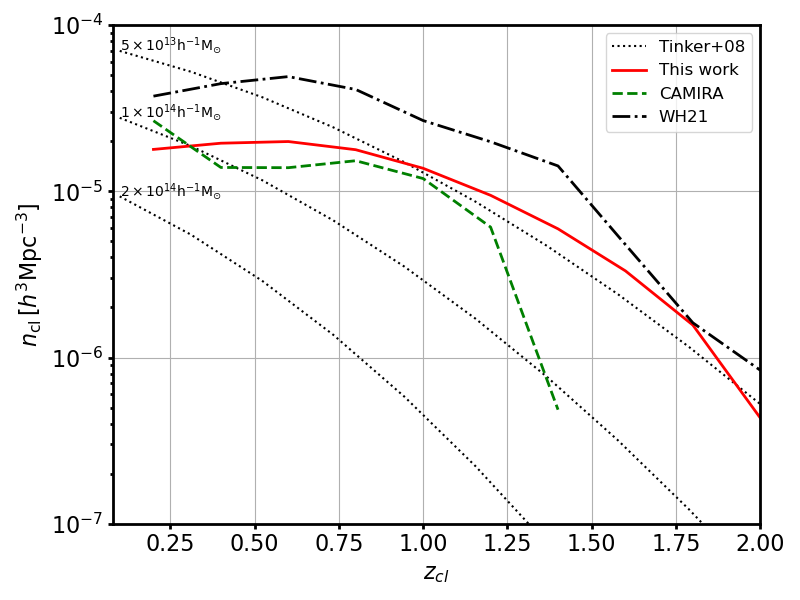}
    \caption{Comoving number density of galaxy structure candidates as a function of cluster photometric redshift. The dotted lines represent the halo mass functions obtained by \citet{tinker08} for three different mass limits, as indicated in the figure. The solid red, dashed green, and dash-dotted black lines denote the results obtained by this work, CAMIRA PDR3, and WH21, respectively.   
}
    \label{fig: cl_n_dens}
\end{figure}

Figure \ref{fig: cl_n_dens} shows the comoving number density of the selected structures as a function of their photometric redshift, determined as the average photometric redshift of the member galaxies ($z_{cl}$). For comparison, we include the expected number densities derived from the halo mass function of \citet{tinker08}, assuming the same cosmology adopted in this work. The theoretical curves correspond to halos with constant mass thresholds, as indicated in the figure. These theoretical curves serve as a reference to evaluate the redshift evolution of the detected structures and to compare the results obtained by different cluster-finding algorithms, helping to assess their completeness. The solid red, dashed green, and dash-dotted black lines denote the results obtained by this work, CAMIRA, and WH21, respectively. Note that CAMIRA selected 20,654 candidates over an effective area of $\rm \sim 950 \ deg^{2}$ of the HSC-SSP Wide PDR3, while WH21 selected 21,661 candidates in an effective area of $\rm \sim 800 \ deg^{2}$.

Up to $\rm z \sim 1.25$, our selection agrees well with that obtained by CAMIRA. WH21 selects a considerably larger number of candidates proportionally to its survey area. The curves decrease more gradually than the theoretical models for fixed upper mass limits until $z \sim 1$. In our case, within the range $\rm 1 < z < 2$, the curve closely follows the trend of the theoretical curve for $\rm M < 5 \times 10^{13} \ h^{-1}M_{\odot}$. A similar trend is observed with the CAMIRA selection, but its curve starts to decline more quickly around $\rm z \sim 1.2$, near the limit of their selection range.

The elevated number density in WH21 likely indicates that their selection captures a larger number of lower-mass galaxy groups, some of which may not evolve into massive clusters. This is supported by the fact that their number density follows a trend that lies well above the expected theoretical predictions for $\rm M_{halo} = 10^{14} \ M_{\odot}$ halos, suggesting that their catalog includes a larger fraction of structures residing in lower-mass halos compared to the other catalogs. While this may enhance the completeness of their selection, it could also introduce a higher fraction of contamination from field galaxy overdensities that do not correspond to actual galaxy clusters or protoclusters.

These results are consistent with our analysis using the mocks, where, as we move to higher redshifts, the fraction of structures with masses below $\rm 10^{14} \ M_{\odot}$ - i.e., protoclusters or massive galaxy groups, according to the definition used—also increases (see V25a Figures 12 and 13). While our methodology aims to select systems likely to evolve into clusters by $\rm z = 0$, some contamination from massive groups is expected. Distinguishing between groups and protoclusters in observations remains challenging, but future simulation-based studies may provide observable criteria to aid in this separation.

\begin{figure}

	\includegraphics[width=\columnwidth]{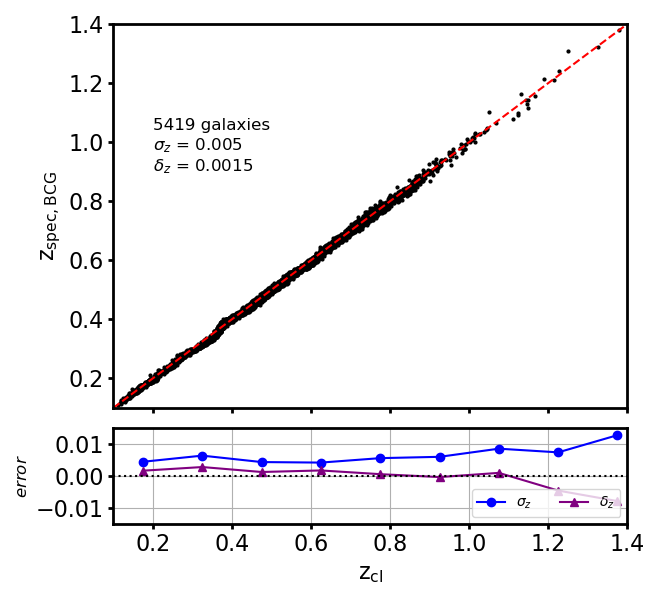}
    \caption{Upper panel: Spectroscopic redshift of the dominant galaxies ($\rm z_{spec, BCG}$) as a function of the redshift of the structure they inhabit ($\rm z_{cl}$). Lower panel: Metrics as a function of $\rm z_{cl}$. The blue line represents the dispersion $\sigma_{z}$, while the purple line indicates the bias $\delta_{z}$.  
}
    \label{fig: pz_perf}
\end{figure}

To evaluate the performance of the photometric redshifts, we compared the spectroscopic redshift of dominant galaxies ($\rm z_{spec, BCG}$) that have such measurements with the redshift of the structure, as described in Section \ref{sec: review} and, in more detail for the membership determination, in V25a. The results are shown in Figure \ref{fig: pz_perf}. We consider two metrics to assess the results: one for random deviations, $\sigma_{z}$, and another for systematic deviations, $\delta_{z}$, which are calculated as the standard deviation and the mean of $\rm (z_{cl} - z_{spec, BCG})/(1 + z_{spec,BCG})$, respectively. There is an excellent agreement between photometric and spectroscopic redshifts, with global metrics of $\sigma_{z} = 0.005$ and $\delta_{z} = 0.0015$, similar to the values found by \citet{oguri18}. In the bottom panel of Figure \ref{fig: pz_perf}, we show how these metrics vary as a function of $\rm z_{cl}$. Note that even at high redshifts, $\rm z_{cl} > 1.2$, the redshifts remain quite similar.

\subsection{Comparisons with CAMIRA and WH21 cluster candidates}
\label{sec: cfs_comp}

\subsubsection{Cross-match fractions}
\label{sec: xmatch}

We performed a cross-match analysis of our catalog with CAMIRA and WH21 within different redshift intervals. A cluster is considered to have a cross-match in the other catalog if it falls within the same redshift slice, i.e., $\rm [z_{cl} \pm 0.05 \times (1 + z_{cl})]$–-where the factor 0.05 accounts for the typical photometric redshift uncertainty in the HSC-SSP Wide sample \citep{tanaka18, nishizawa20}–-and the BCGs are within an angular diameter transverse distance of 1 Mpc.

\begin{figure*}

	\includegraphics[width=\textwidth]{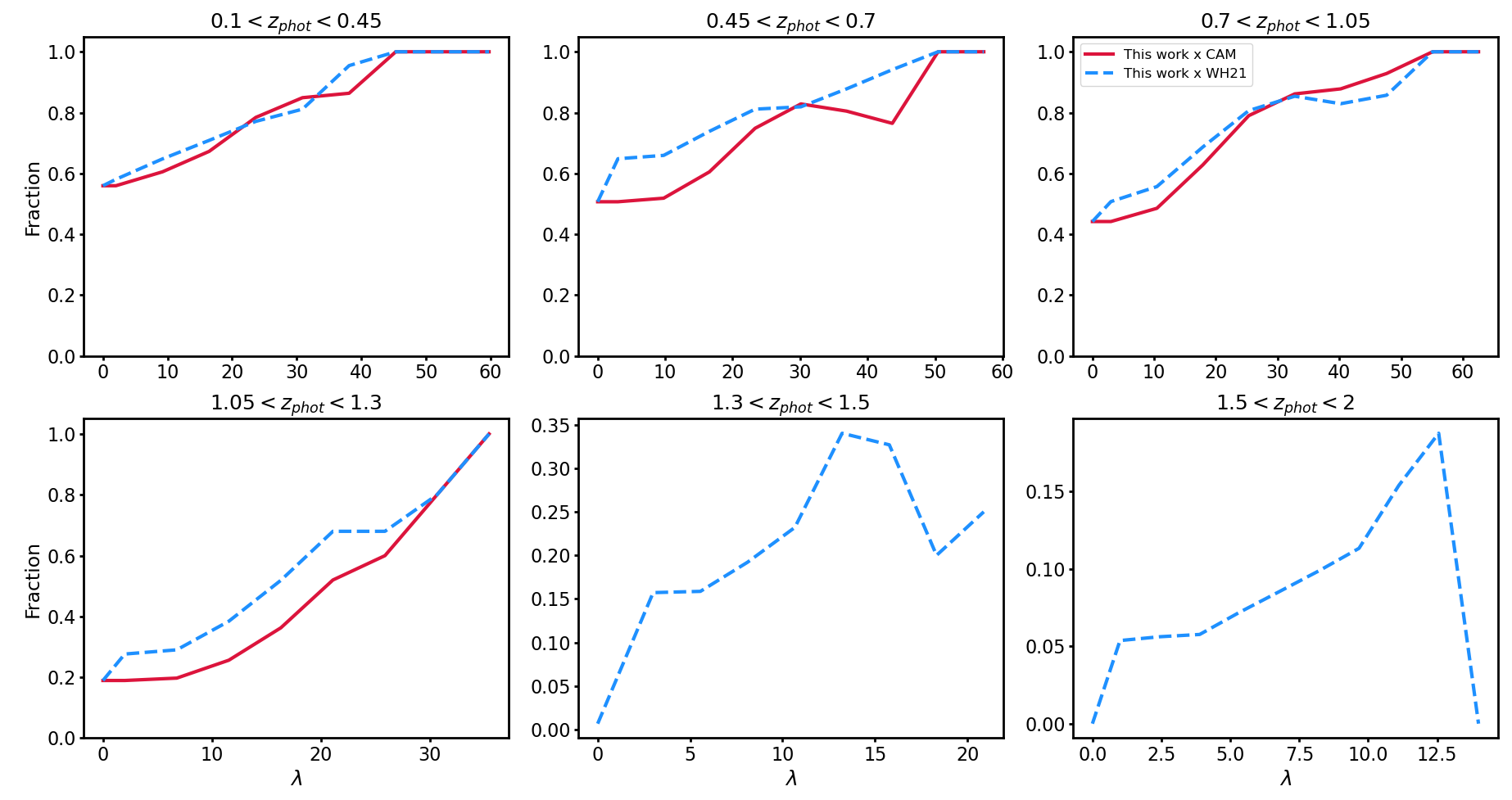}
    \caption{Fraction of cross-matched structures as a function of richness. Each plot shows the results in a given redshift interval, as indicated above each panel. Solid red and dashed blue denote the comparison between this work with CAMIRA and WH21, respectively. The fraction is calculated by applying cuts on the estimated richness from this work, defined as the ratio between the number of structures in our catalog with richness above a given threshold that have a match in the other catalogs (solid red and dashed blue lines for CAMIRA and WH21, respectively), and the total number of structures in our catalog above that same threshold. Therefore, a richness cut of zero corresponds to considering the entire sample.
}
    \label{fig: xmatch}
\end{figure*}

Figure \ref{fig: xmatch} shows the fractions of clusters in common in six redshift ranges as a function of richness. The fraction is calculated by applying cuts on the estimated richness from this work, defined as the ratio between the number of structures in our catalog with richness above a given threshold that have a match in the other catalogs (solid red and dashed blue lines for CAMIRA and WH21, respectively), and the total number of structures in our catalog above that same threshold. Therefore, a richness cut of zero corresponds to considering the entire sample.

As expected, there is an increase in the fraction of structures with a cross-match as richness increases. Notably, both curves show a similar trend, indicating that, regardless of the catalog our selection is compared against, the fraction of matching structures as a function of richness and redshift remains consistent. However, this consistency does not imply that the exact same structures are being detected across the three catalogs. At high redshifts, $\rm 1.5 < z < 2$, the fraction of candidates matching between our selection and WH21 is quite low ($\lesssim$ 20\%; see Section \ref{sec: match_unmatch}). A significant distinction to highlight is that WH21 limits their selection to galaxies with a cross-match in the unWISE catalog \citep{schlafly18}, a restriction not applied by us or by CAMIRA.

\begin{figure*}

    \centering
	\includegraphics[width=0.93\textwidth]{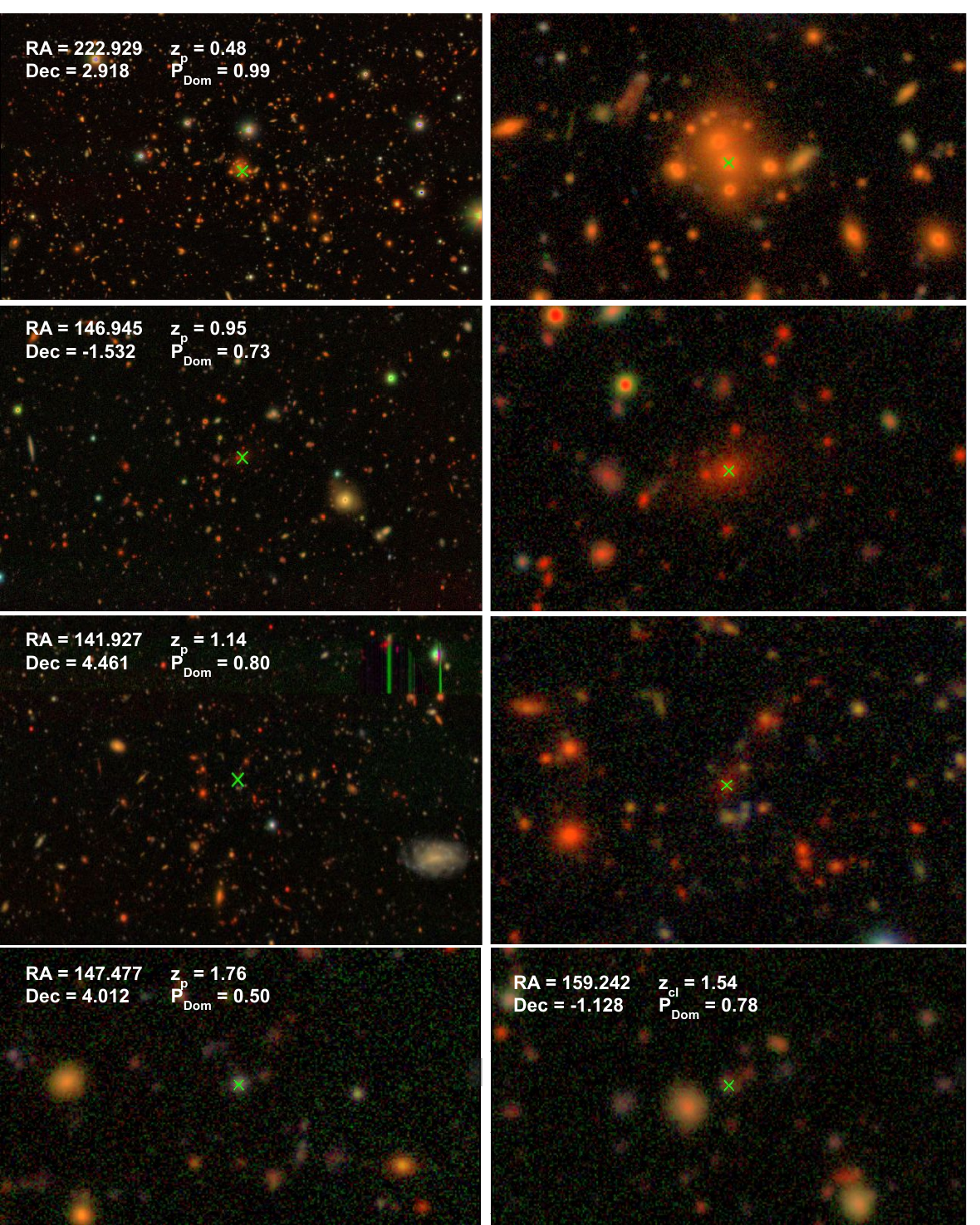}
    \caption{Examples of (proto)clusters of galaxies selected in the catalog produced by this work that have no counterparts in the catalogs generated by CAMIRA and WH21. The first three rows images display one structure per row. On the left side, a wider view of approximately $\sim 0.8$ Mpc per side is shown, while on the right, a fourfold zoom-in is provided. In the last row, we present two high-redshift examples with only the zoomed-in views. The position (indicated by a green `x'), photometric redshift, and the probability of the dominant galaxy are marked on each image.
}
    \label{fig: nomatch_examples}
\end{figure*}

These diagrams clearly illustrate that different methods for selecting galaxy cluster candidates yield varied results, leaving room for unique discoveries. Figure \ref{fig: nomatch_examples} provides RGB images from the $g$, $r$, and $i$ bands, showcasing examples of candidates identified in this study that lack cross-matches in the other catalogs.

\subsection{Matched and non-matched candidates}
\label{sec: match_unmatch}

\begin{figure}

	\includegraphics[width=\columnwidth]{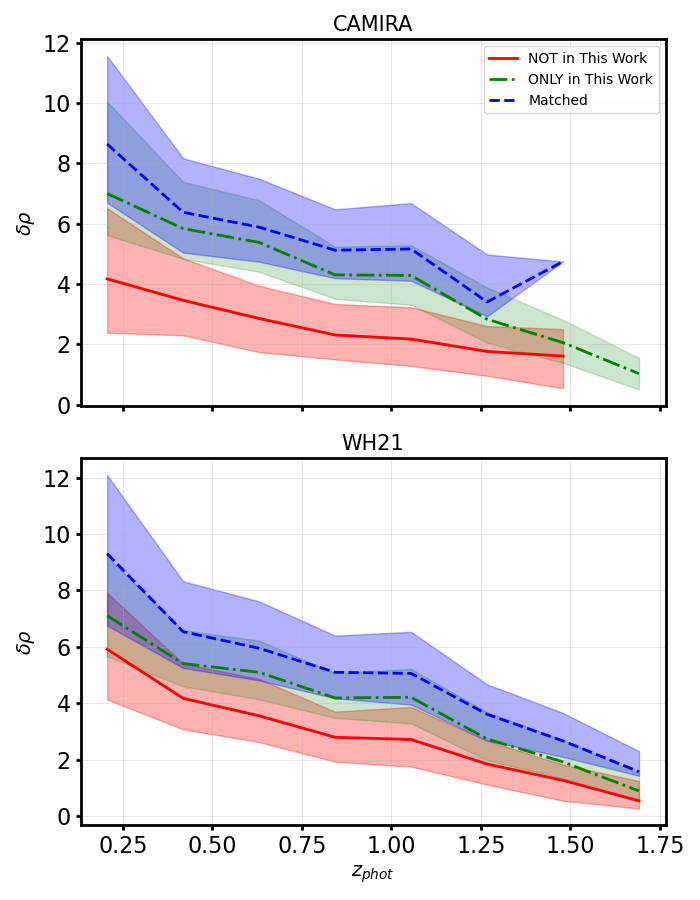}
    \caption{Median stellar mass density contrast ($\delta \rho$) as a function of photometric redshift for cluster candidates in CAMIRA (top) and WH21 (bottom). In each panel, dashed blue curves represent matched systems between our catalog and the external one. Solid red curves indicate systems found only in the external catalog, while dash-dotted green curves represent candidates identified only in our catalog. Shaded regions correspond to the interquartile range in each redshift bin.  
}
    \label{fig: density_cont_mat_nonmat}
\end{figure}

To investigate the nature of the structures identified by CAMIRA and WH21 that are not recovered by our algorithm or those structures that are only in our selection, we divided the candidates from each catalog into three categories: those matched with our catalog, those found only in the external catalog, and those found only in our work. For each case, we measured the local stellar mass density contrast ($\delta \rho$) around the candidate BCGs using the methodology described in V25a. As shown in Figure \ref{fig: density_cont_mat_nonmat}, matched structures systematically exhibit the highest density contrasts across the redshift range, while systems missed by our algorithm tend to show significantly lower $\delta \rho$. This indicates that, despite being identified in CAMIRA or WH21, many of these structures do not present strong enough overdensities under our definition. Visual inspection further supports this, as most of the non-matched candidates lack a clearly concentrated galaxy distribution.

Interestingly, candidates identified only by our algorithm—i.e., not present in CAMIRA or WH21—tend to show intermediate density contrast values, generally higher than the missed systems, though slightly lower than the matched ones. These results reinforce that the selection of galaxy clusters is highly sensitive to the adopted methodology. They also highlight that our algorithm is effectively identifying new structures with significant density contrast, even if they are not captured by other approaches.

\begin{figure}

	\includegraphics[width=\columnwidth]{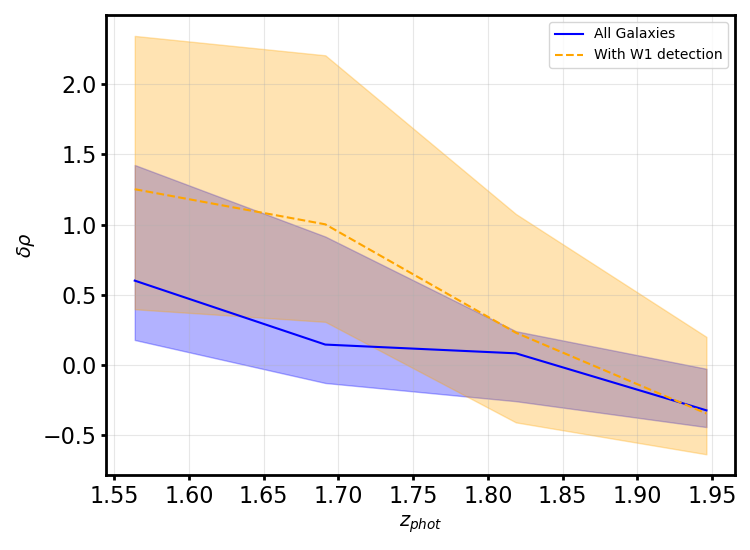}
    \caption{Stellar mass density contrast ($\delta\rho$) as a function of photometric redshift for WH21 unmatched candidates at $\rm z > 1.5$. The solid blue line shows the results using all galaxies in our dataset, while the dashed orange line shows the contrast computed using only galaxies with photometric measurements in the W1 band. Shaded areas represent the inter-quartile interval.  
}
    \label{fig: wh21_highz_nonmat}
\end{figure}

This trend becomes especially relevant at high redshifts, where the matching fraction between our candidates and WH21 drops considerably—below 20\% for $\rm z_{phot} > 1.5$. To further explore the nature of these unmatched WH21 candidates, we estimated the density contrast around their coordinates using our full galaxy sample and selection criteria. Additionally, we repeated the calculation using only galaxies with W1-band photometric detections, as required by WH21. The results, presented in Figure \ref{fig: wh21_highz_nonmat}, indicate that the measured overdensities are systematically higher when restricted to W1-detected galaxies. This highlights an important distinction: the lower matching rate at high redshift arises not necessarily because of spurious detections, but due to a combination of methodological differences and sample selection effects. Specifically, by relying on a subsample limited to galaxies with W1-band detections, the WH21 algorithm traces overdensities with a different galaxy population—one that does not fully overlap with the more complete stellar mass-selected sample used in our approach.

\subsubsection{Structure`s richness and redshift}
\label{sec: rich_and_z_comp}

\begin{figure*}

	\includegraphics[width=\textwidth]{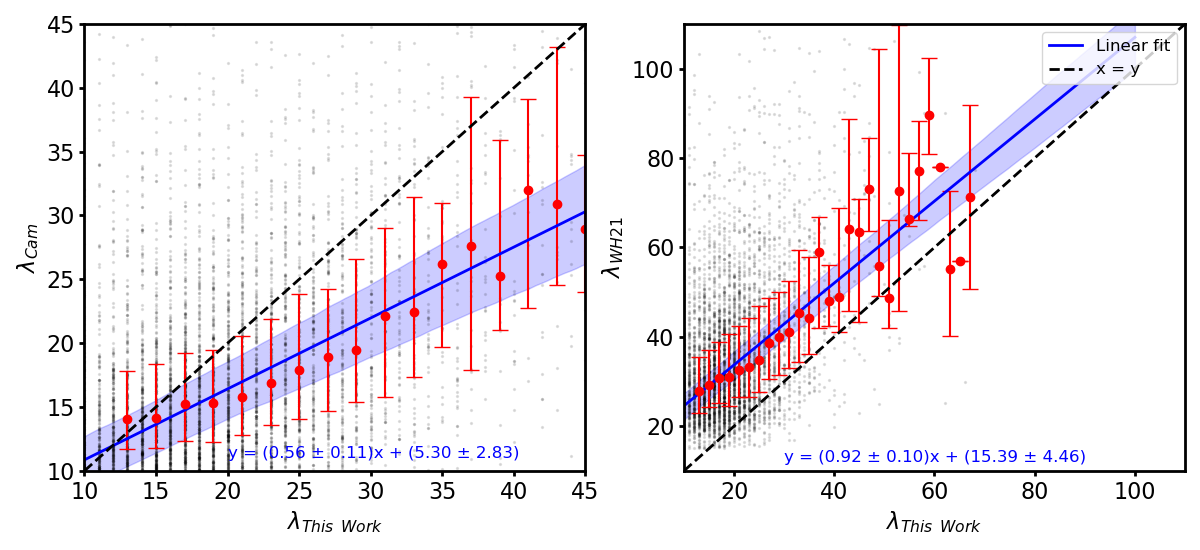}
 	\includegraphics[width=\textwidth]{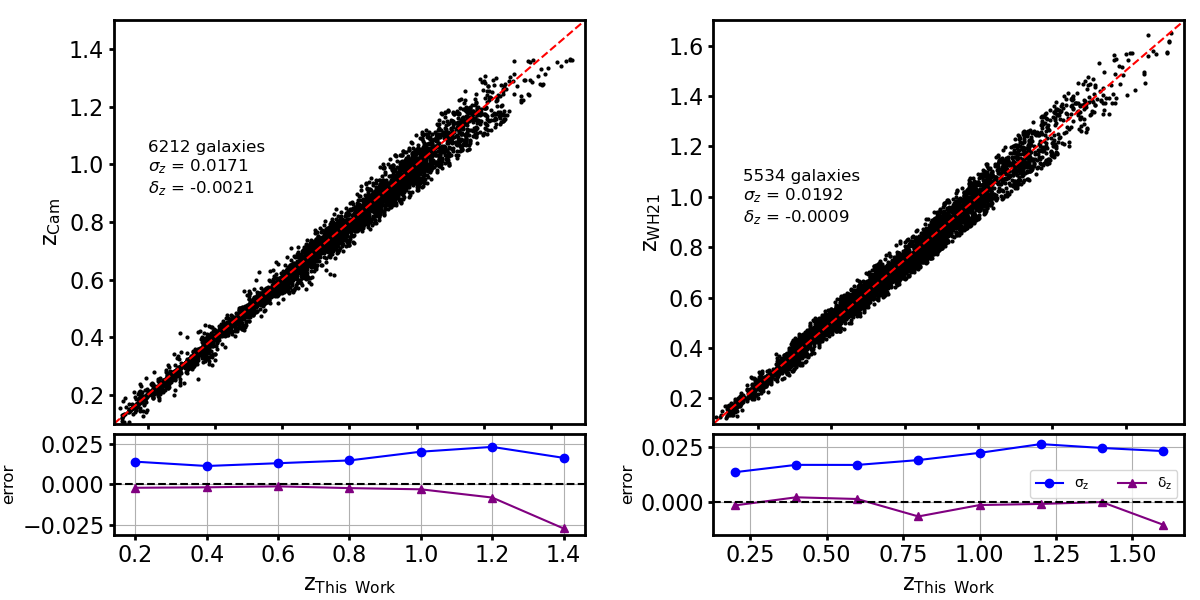}

    \caption{Upper plots: structure's richness comparison with CAMIRA (left) and WH21 (right). Red dots denote the median and error bars delimit the interquartile range. Dashed (solid) blue line denotes x = y (linear fit). Lower plots: Structure's photometric redshift comparison. The purple and blue lines represent the bias and RMS error, respectively, as functions of redshift, displayed beneath each panel.
}
    \label{fig: rich_pz_comparison}
\end{figure*}

In Figure \ref{fig: rich_pz_comparison}, we present a comparison between the richness (see Section \ref{sec: review}) and the photometric redshifts of the structures identified in this study with those from CAMIRA and WH21. Although all three methods rely on similar general principles—such as selecting member galaxies based on their projected distance from a central galaxy and stellar mass thresholds—the specific criteria adopted in each approach differ. These differences include how the central galaxy is defined, the stellar mass limits applied to member selection, and the radial distance thresholds used to define cluster membership. As a result, variations in the richness assigned to the same structures are expected. Our method, for instance, yields richness values systematically higher than CAMIRA, while lower than WH21. 

In the lower plots, comparing the photometric redshifts, there is an excellent correlation in both cases. The dispersion, measured by the root mean square error, reaches a maximum of approximately 0.025 around $\rm z = 1.2 - 1.4$.

\subsubsection{BCG candidates offset}

\begin{figure*}

	\includegraphics[width=\textwidth]{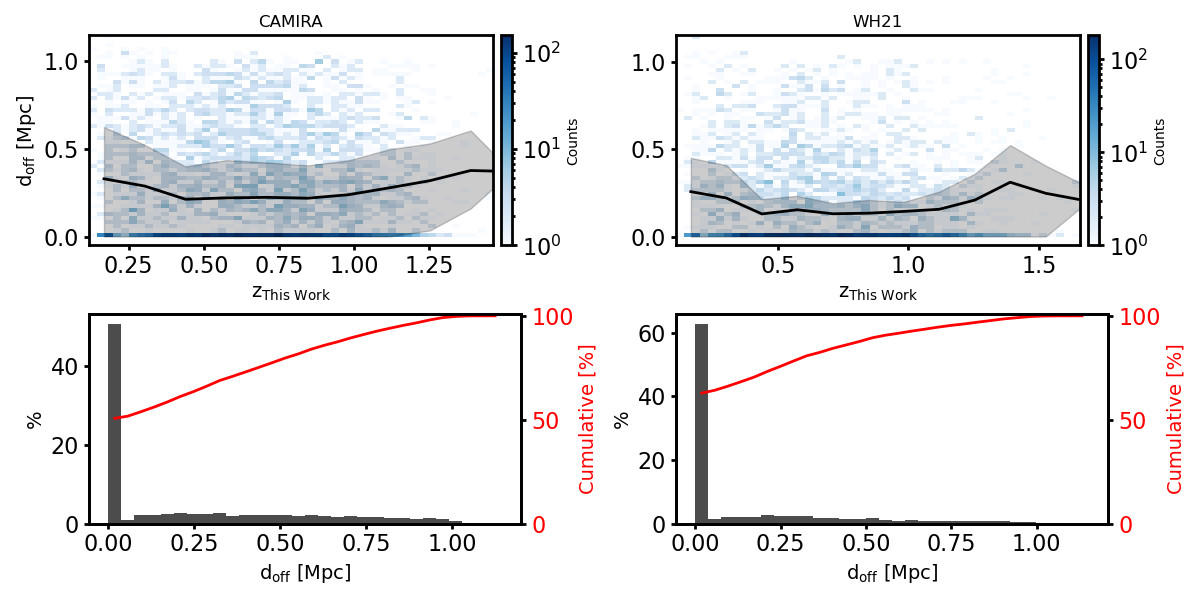}

    \caption{Projected offset between the position of the BCGs identified in this work and those from the CAMIRA (left) and WH21 (right) cluster catalogs. The top panels show the distribution of projected offsets ($d_{\rm off}$) as a function of redshift, with the density of points represented in logarithmic scale. Black lines denote the mean and the shaded area encompasses the inter-quartile range. The bottom panels show the normalized histograms of the offsets in units of percentage and the cumulative fraction of clusters as a function of the BCG offset is shown as a red line, with values referenced to the right y-axis. While $d_{\rm off}$ = 0 reflects cases where both algorithms identify the same galaxy as the BCG, the rising trend up to $\sim$ 0.25 Mpc and subsequent decline likely reflects the typical scale of cluster cores, where different bright galaxies can be selected as BCGs depending on the method.
}
    \label{fig: offset_comparison}
\end{figure*}

To further investigate the consistency of the BCG identification across catalogs, we examine the projected offsets between the central galaxies identified in this work and those reported in CAMIRA and WH21 for matched clusters. As shown in Figure \ref{fig: offset_comparison}—which displays the distribution of projected offsets (lower panels) and their evolution with redshift (upper panels)—a significant fraction of the matched systems corresponds to exact matches in BCG identification: approximately 50\% for CAMIRA and 62\% for WH21. Among the remaining systems, 62\% (CAMIRA) and 75\% (WH21) are lying within 0.25 Mpc, and over 90\% within 0.75 Mpc. There is no evident correlation between the BCG offset and redshift, suggesting that the agreement in central galaxy identification is similar across the redshift range probed in this work.

A more detailed comparison between the BCG positions and the centers of X-ray-detected clusters is presented in Section \ref{sec: xray_offset}.

\subsection{High-z candidates}
\label{sec: highz_samp}

\begin{figure*}

	\includegraphics[width=\textwidth]{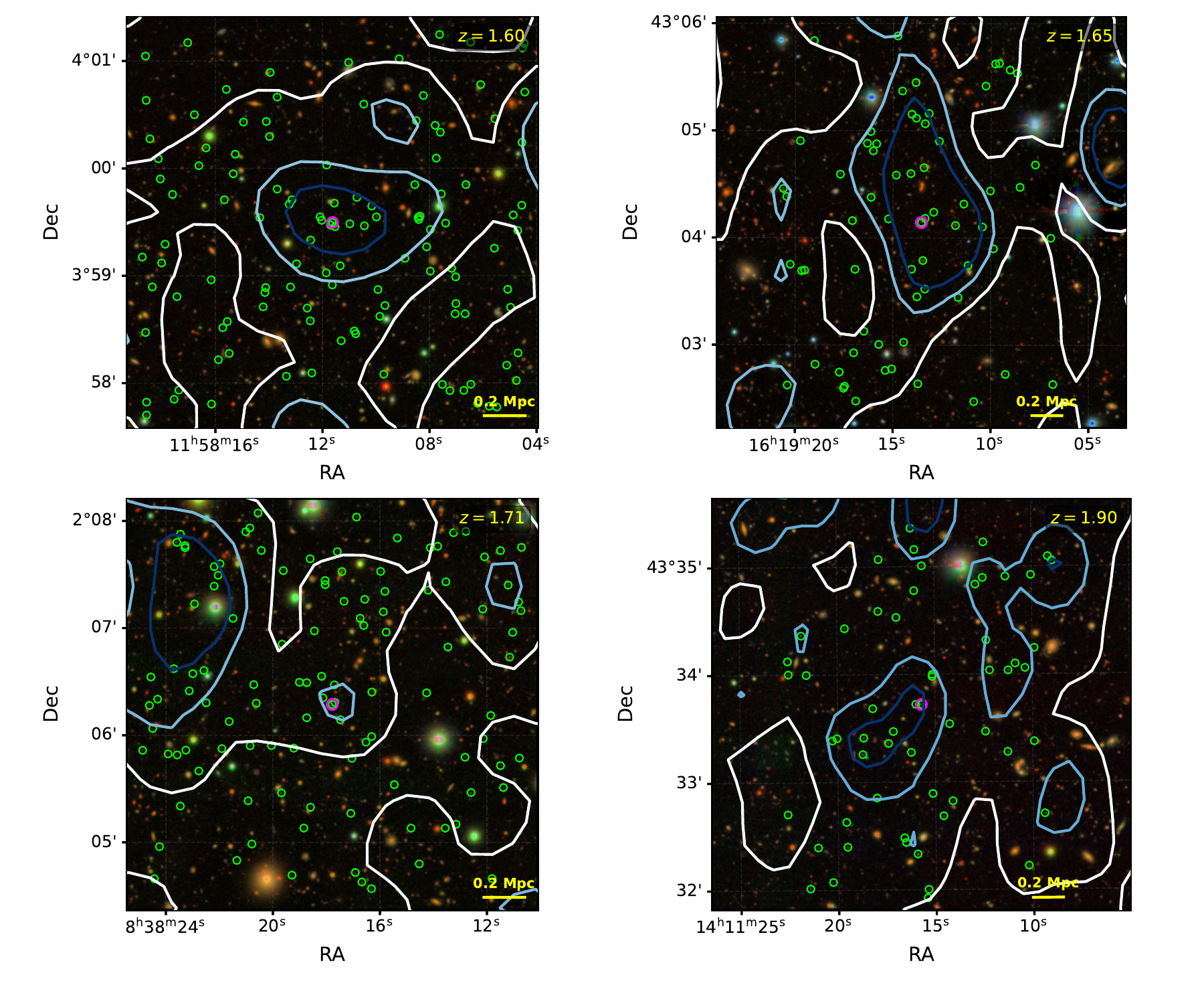}

    \caption{RGB images of four high-redshift cluster candidates ($\rm 1.5 < z_{phot} < 2.0$) identified by our method. Each panel shows a $\rm 1 \ Mpc \times 1 \ Mpc$ region centered on the (proto)BCG candidate, marked with a magenta circle. Galaxies within the photometric redshift slice of the structure are highlighted in green. Contours trace the 1, 2, and 3$\sigma$ (white-to-blue) projected stellar mass density, calculated from the selected galaxies in the redshift slice and smoothed with a Gaussian filter. The scale bar in each panel corresponds to 0.2 Mpc at the redshift of the structure, indicated in the top-right corner. 
}
    \label{fig: highz_cand}
\end{figure*}

To assess the reliability of our high-redshift cluster candidates ($\rm 1.5 < z_{phot} < 2.0$), we performed a more detailed analysis of their local environments. Figure \ref{fig: highz_cand} presents RGB images of four representative candidates, overlaid with contours of stellar mass density contrast computed using the same selection criteria adopted in our method. 

These contours indicate regions at 1, 2, and 3$\sigma$ above the mean field density, computed within the photometric redshift slice of each structure by applying a Gaussian filter to a 2D histogram weighted by stellar mass. These visualizations show that the BCGs (highlighted in magenta) are typically located near overdensity peak, often within the 3$\sigma$ contours, reinforcing that these systems trace significant mass concentrations. 

Although visual inspection alone becomes less conclusive at these redshifts due to the limitations of imaging, our analysis demonstrates that these candidates are not isolated or spurious detections, but rather are embedded in statistically significant overdense environments within the uncertainties of the photometric redshifts. Additionally, this observational result is consistent with our mock-based analysis (V25a), which shows a high incidence of (proto)cluster environments at these redshifts. 

%The density maps presented here were computed independently from our cluster-finding algorithm, further validating the robustness of our selection through an alternative method sensitive to the same physical features we aim to detect. 

\begin{figure*}
    \centering
	\includegraphics[width=0.9\textwidth]{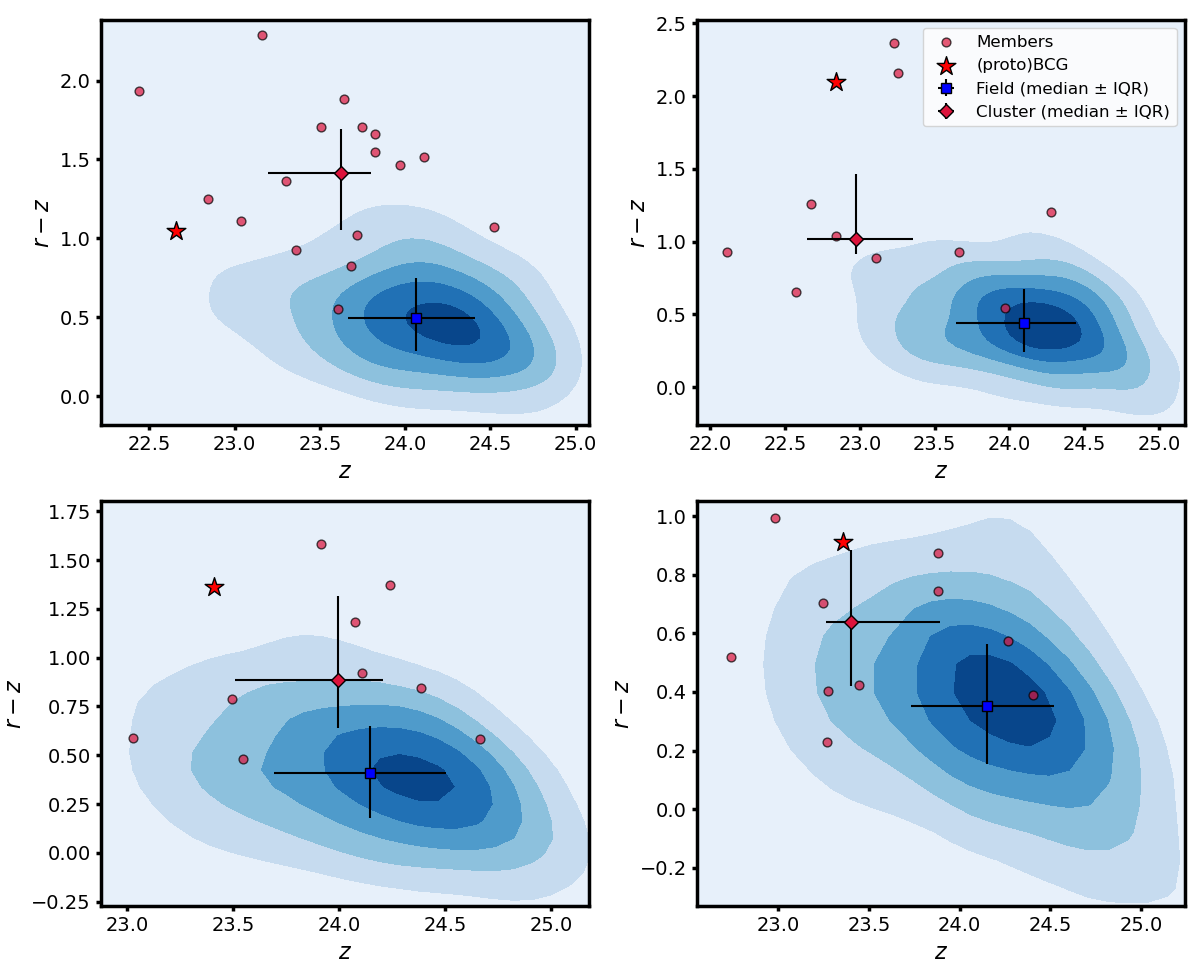}
	\includegraphics[width=0.5\textwidth]{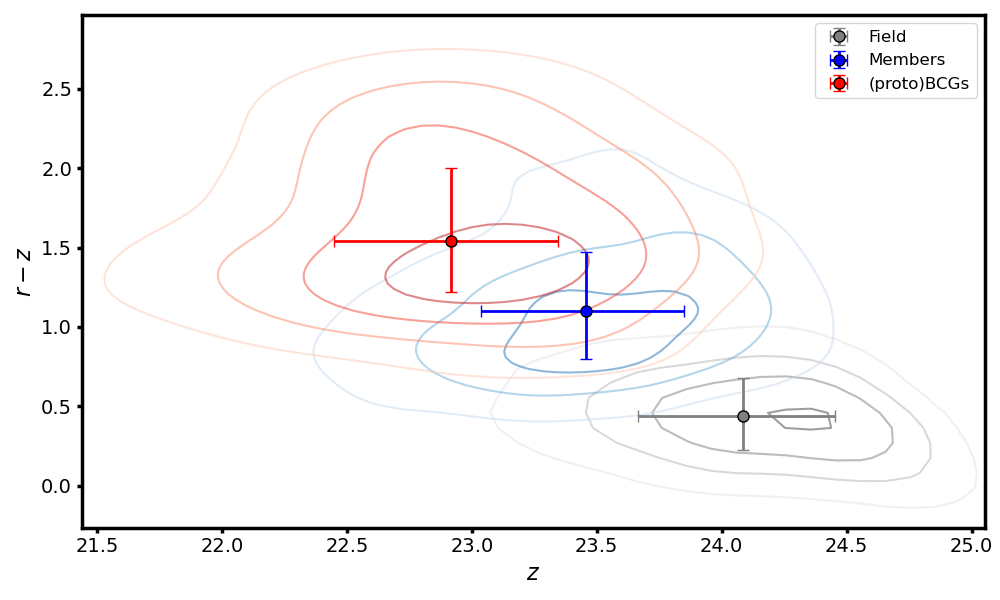}

    \caption{Color–magnitude diagrams ($r-z$ vs. $z$) for high-redshift ($\rm 1.5 < z_{phot} < 2.0$) (proto)cluster candidates selected in this work. Top panels: Individual diagrams for the four examples shown in Figure \ref{fig: highz_cand}. Red circles stand for member galaxies, red stars are (proto)BCGs, and the underlying field galaxy population, at the same photometric redshift slice, as blue KDE contours. Error bars indicate the medians and interquartile ranges (IQR) of color and magnitude for the cluster members and the field. Bottom panel: Combined color–magnitude diagram for all structures with $\rm z_{phot} > 1.5$. The field galaxy population (gray contours), member galaxies (blue), and (proto)BCGs (red) are shown as KDE contours. Markers indicate the median and IQR for each population.
}
    \label{fig: highz_cand_colmag}
\end{figure*}

Additionally, we analyze their photometric properties in observed color–magnitude space. Figure~\ref{fig: highz_cand_colmag} presents $r-z$ versus $z$ diagrams for the four (proto)clusters examples shown in Figure \ref{fig: highz_cand} in the redshift range $\rm 1.5 < z_{phot} < 2.0$, as well as a combined diagram stacking all (proto)cluster candidates above $\rm z_{phot} > 1.5$ (bottom panel). In each case, member galaxies and the central (proto)BCG are contrasted with the surrounding field population at similar redshifts. 

The diagrams reveal a clear separation in color and magnitude between (proto)BCGs, (proto)cluster members, and field galaxies, with (proto)BCGs appearing redder and brighter, followed by cluster members and field galaxies for galaxies. This distinction is especially apparent in the stacked diagram, where the KDE contours and medians emphasize the systematic offset between these galaxy populations. These results strengthen the interpretation that the selected structures correspond to genuine (proto)clusters rather than statistical fluctuations, even in the challenging high-z regime.

Table \ref{tab: appendix} lists key properties of a subset of 25 high-redshift (proto)cluster candidates is provided in Appendix \ref{apx: highz_sample}, including the four examples discussed in this section.

\subsection{(proto)BCGs and cluster members properties}
\label{sec: members_props}

The selection of structures in this work is made by first calculating the probability of a given galaxy being dominant, i.e., identifying (proto)BCGs, and subsequently identifying the satellite galaxies (see Section \ref{sec: review}). In CAMIRA and WH21, the procedure occurs in reverse: potential structures are first identified through concentrations of galaxies, and then the dominant galaxy is determined. In this section, we will compare the distribution of properties of the dominant galaxies as a function of redshift across the different selections.

\begin{figure*}

	\includegraphics[width=\textwidth]{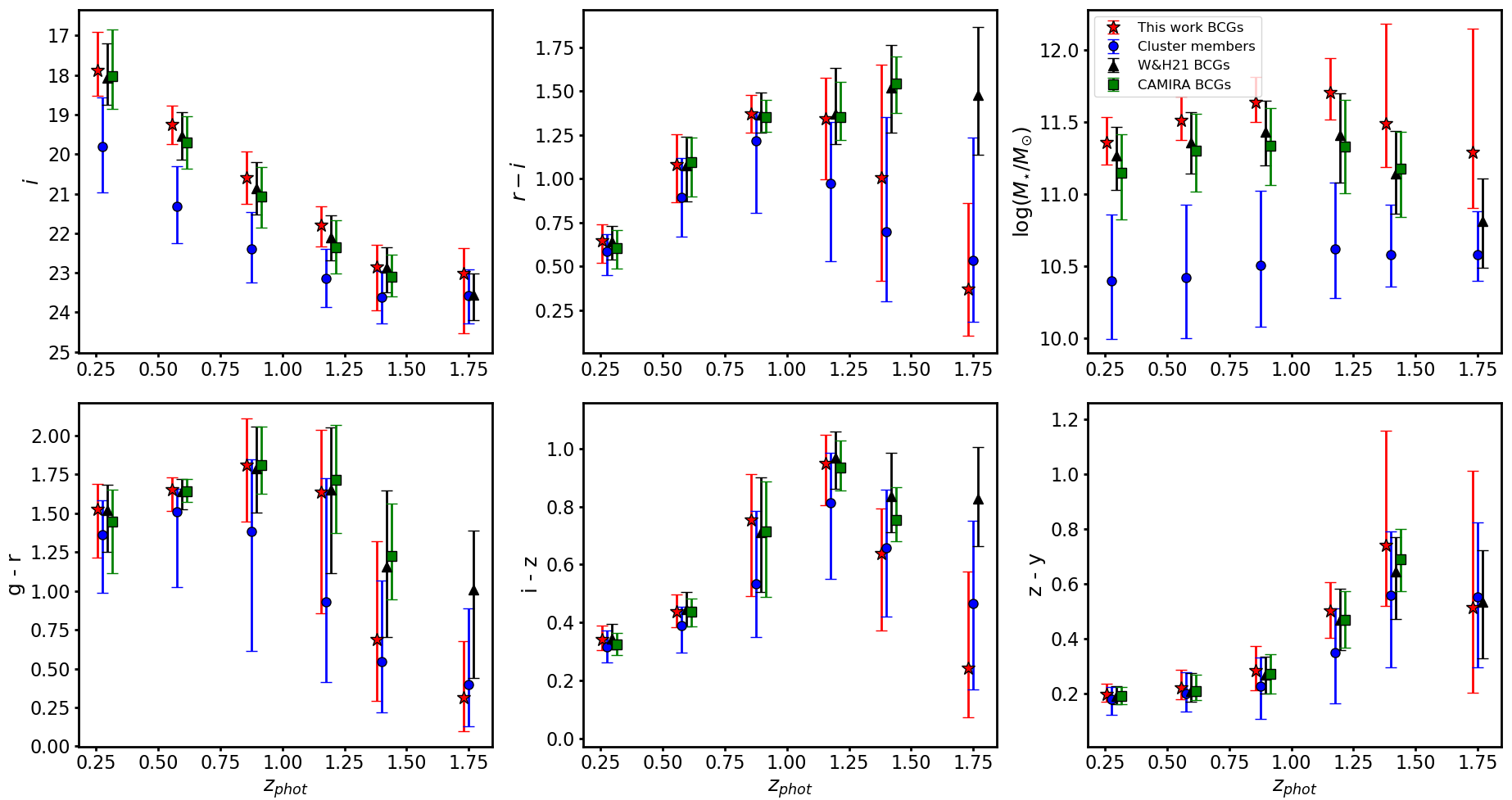}

    \caption{First row, from left to right: BCGs $i$-band magnitude, $r - i$ observed color, and stellar mass as function of photometric redshift. Second row, $g - r$, $i - z$, and $z - y$ observed colors as a function of photometric redshift. Colors are in the observed frame. Points denote the median properties within a given redshift bin, and the error bars are the dispersions bounded by the 16th and 84th percentiles. BCGs from CAMIRA and WH21 catalogs are represented in the plots by green squares and black triangles, respectively, while BCGs from this work are denoted by red stars. Blue dots denote satellite galaxies selected in this work. 
}
    \label{fig: bcg_properties_comp}
\end{figure*}

Figure \ref{fig: bcg_properties_comp} presents six plots. As expected, satellite galaxies are less luminous and less massive, and the difference compared to the dominant galaxy decreases as redshift increases. There is also a trend for these galaxies to be bluer; however, in the lowest redshift bin, they have similar colors to that of the dominant galaxies. The slight decrease in stellar mass at $\rm z \gtrsim 1.2$ is a consequence of the evolving galaxy population.

There are some interesting differences between our selection of (proto)BCGs and those selected by other catalogs. The (proto)BCGs selected in this work tend to be slightly brighter in the $i$-band and they tend to be more massive than the CAMIRA and WH21 selection across the entire redshift range. For all observed colors other than $z - y$, at high redshifts ($\rm z_{phot} > 1.2$), we identify bluer dominant galaxies, which is particularly evident in the last redshift bin presented in the plot, where the (proto)BCGs are bluer than even the satellite galaxies. For $z - y$, the (proto)BCGs selected in this work are consistent with the findings of WH21. According to our analysis with mocks in V25a, our selection at high redshifts shows a high incidence of protoBCGs, i.e., dominant galaxies in protoclusters, which are more associated with higher star formation rates. WH21 selects structures based on concentrations of galaxies that have counterparts in the unWISE mid-infrared catalog. For this reason, significant differences in the selections are likely, especially at high redshifts since their analysis is limited by the sensitivity of the WISE survey.

\subsection{Comparison with X-ray cluster catalogs}
\label{sec: xray}

In this section, we use two catalogs of galaxy clusters detected via extended X-ray emission: the first catalog produced by eROSITA in the Western Galactic Hemisphere \citep{bulbul24} and the second release of the Meta-Catalog of X-ray Detected Clusters of Galaxies \citep[MCXC-II; ][]{sadibekova24}, a compilation of several publicly available X-ray cluster catalogs based on the ROSAT All-Sky Survey and serendipitous detections. 

To conduct the comparisons, we cross-match each catalog of clusters detected in optical surveys — including this work, CAMIRA, and WH21 — with catalogs of clusters detected in X-rays. This will allow us to compare the richness in each of these catalogs with the X-ray estimated mass ($M_{500}$), as well as to evaluate the offset between the position of the dominant galaxy and the centroid in the X-ray cluster catalogs.

\subsubsection{X-ray cluster masses vs. richness}

We preselected entries in these catalogs with an X-ray estimated mass of $\rm \log(M_{X, 500}/M_{\odot}) > 13$, temperature $\rm kT_{X} > 1 \ keV$, and redshift $\rm z > 0.1$. After this selection, the combined total of identified clusters across both catalogs is 3,718. Limiting to the effective area of the HSC-SSP Wide survey considered in this work, we find 93 objects. Of these, 40 (not necessarily the same) have a cross-match with our catalog and the WH21 catalog, and 57 with the catalog produced by CAMIRA. The number of matches when we limit the samples to $\rm z \geq 0.2$ is 29, 19, and 28 for our catalog, WH21, and CAMIRA, respectively. This suggests that our sample may be incomplete in the $\rm 0.1 \leq z_{phot} \leq 0.2$ range. To better understand why we are not identifying these structures at low redshifts, we cross-referenced the CAMIRA catalog coordinates to identify galaxies within the same redshift slice and a radius of 1 Mpc. In over 90\% of cases, there were no galaxies massive enough to meet our pre-selection criterion for dominant galaxies at these redshifts of $\rm M_{\star} \geq 10^{11} \ M_{\odot}$, as described in V25a.

For the structures in the X-ray catalogs that do not match with ours, we used their provided coordinates to determine the BCG as the most massive galaxy within a radius of 1 Mpc, and we identified members according to the same criteria described in this work. We refer to this procedure as a forced selection, since it applies our selection methodology centered on external coordinates rather than on our own detection. Most of the BCG candidates identified through this approach have stellar masses in the range $\rm \log(M_{\star}/M_{\odot}) = 10.5$–-$11$ and, based on visual inspection, appear to be associated with less massive systems, more consistent with galaxy groups.

\begin{figure*}

	\includegraphics[width=\textwidth]{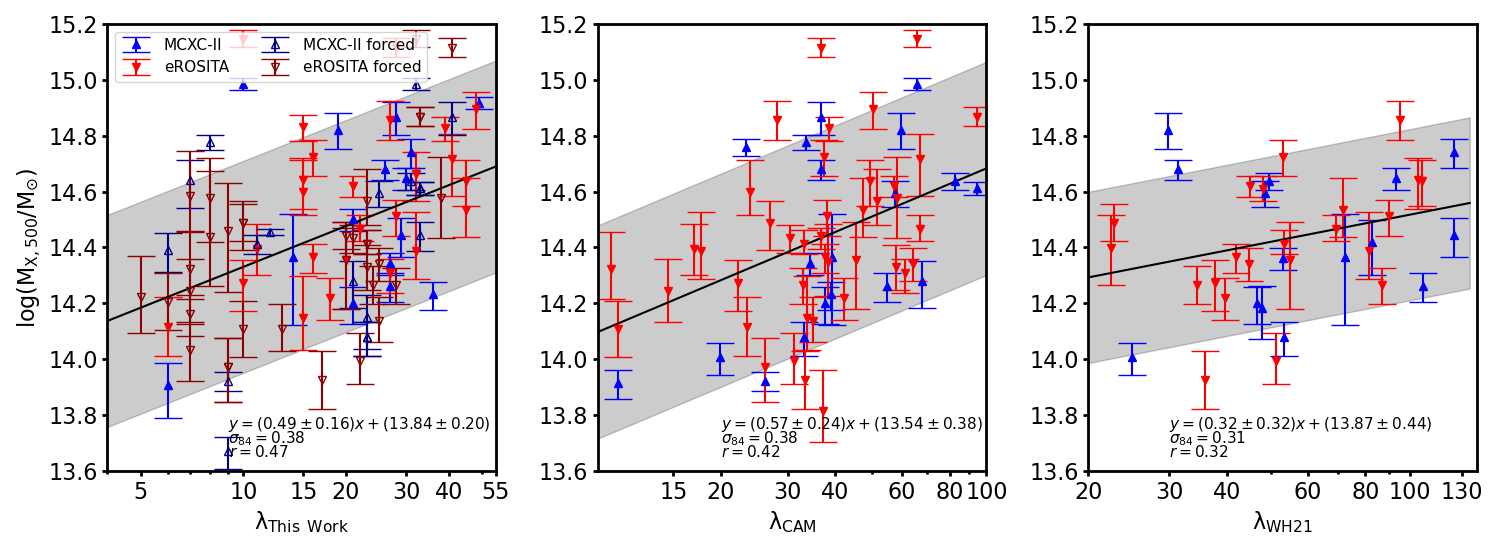}

    \caption{From left to right: Halo masses estimated from X-ray luminosity as a function of richness obtained in this work, CAMIRA, and WH21, respectively. Red filled inverted triangles represent objects matched with the eROSITA catalog, while blue filled triangles indicate matches with the MCXC-II catalog. In the left panel, richness was computed by forcing the detection of BCGs and their members using the coordinates from the X-ray catalogs within the overlapping area of the HSC-SSP Wide survey. These cases are indicated by dark red (blue) unfilled markers for eROSITA and MCXC-II catalog. At the bottom of the plots, we include the best-fit function parameters with their uncertainties, along with $\sigma_{84}$, representing the 84\% confidence interval, and the Pearson correlation coefficients ($r$).
}
    \label{fig: xray_mass_richness}
\end{figure*}

Figure \ref{fig: xray_mass_richness} presents three plots of the X-ray-estimated mass as a function of richness, obtained using the criteria from this work, CAMIRA, and WH21, respectively from left to right. Filled markers represent structures that have a match and, therefore, appear in both the X-ray and HSC-SSP Wide optical catalogs, while unfilled markers represent those found only in the X-ray catalog, for which we performed forced richness estimates. The majority of detections with richness below 10 (90\%), totaling 16 structures, did not match with our catalog. The fitted relation in log-log space (black solid line) is shown at the bottom of each plot. In all three cases, there is a moderate positive correlation between X-ray-derived mass and estimated richness, as indicated by the Pearson correlation coefficient ($r$), which is 0.47, 0.42, and 0.32 from left to right.

\subsubsection{X-ray cluster masses vs. dark matter halo masses}

In V25a, we derived halo mass-richness relations using the dark matter halo masses provided by the Millennium simulation. To obtain these relations, we applied the same selection criteria used in this study to identify member galaxies of structures in mocks that emulate HSC-SSP Wide observations. For more details, see Section 5.2 of V25a.

\begin{figure}

	\includegraphics[width=\columnwidth]{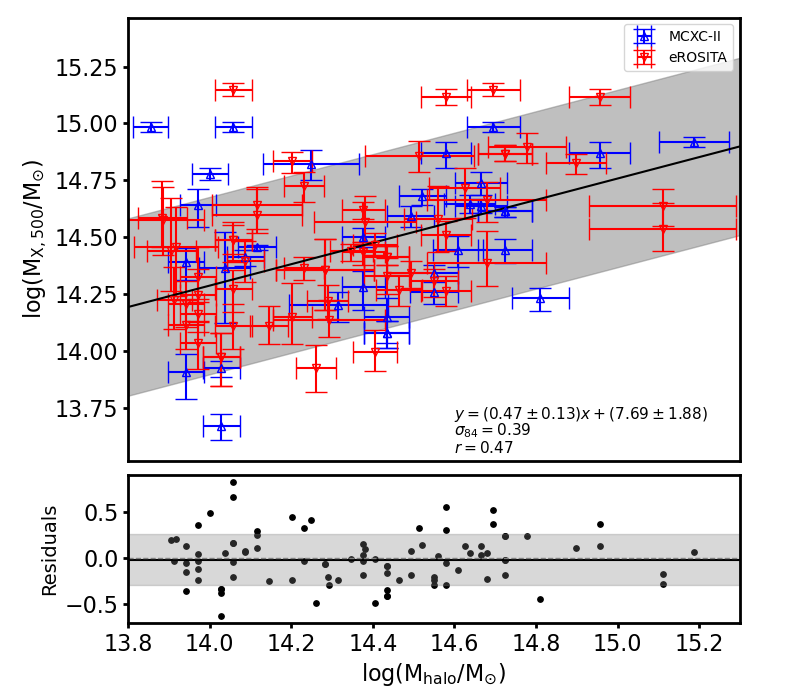}

    \caption{Upper Plot: The mass of the structures derived from X-ray emission as a function of halo mass obtained through mass–richness scaling relations. Red points represent structures from the eROSITA catalog, while blue points correspond to those from the MCXC-II catalog. The black line represents the best-fit function, with the parameters shown below the plot, along with $\sigma_{84}$, representing the 84\% confidence interval, and the Pearson correlation coefficient. Lower Plot: Residuals as a function of halo mass. The horizontal black line indicates the median, and the shaded area represents one standard deviation.
}
    \label{fig: halo_mass}
\end{figure}

Figure \ref{fig: halo_mass} illustrates the relationship between the structure masses obtained from X-ray emission and the dark matter halo masses ($\rm M_{halo}$) derived from the mass-richness relations. As expected, the slope, $\sigma_{84}$, and the correlation coefficient are similar to those obtained for the $\log(M_{X, 500}/M_{\odot})$-$\lambda_{This \ Work}$ relation (left panel of Figure \ref{fig: xray_mass_richness}), as indicated directly in the plot. The residuals are shown in the lower panel. The median of the residuals, represented by the gray line, is zero, while the standard deviation is 0.3 dex, marked by the shaded area.

\subsubsection{Offset between BCG and X-ray cluster centroid}
\label{sec: xray_offset}

The offset between the BCG and the peak of the ICM gas emission carries information about the dynamical state of galaxy clusters. Relaxed clusters, which are not undergoing major interactions or mergers, typically exhibit smaller offsets, reflecting the central position of the BCG within the gravitational potential well. Larger offsets, on the other hand, can indicate ongoing interactions, mergers, or disturbed dynamical states. However, it is also essential to consider that the offset may be influenced by the accuracy of BCG identification in structure detection algorithms. Misidentifications or biases in selecting the dominant galaxy can artificially increase the measured offset \citep[e.g.,][]{ding24}.

\begin{figure}

	\includegraphics[width=\columnwidth]{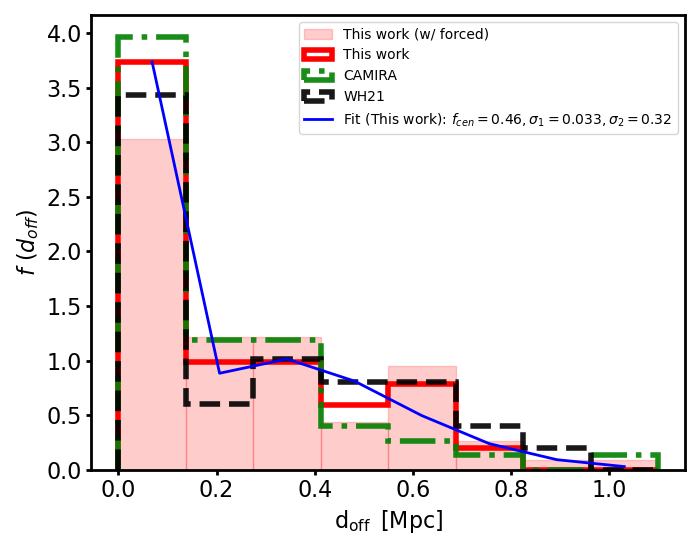}

    \caption{Distribution of the offset between the positions of candidate BCG galaxies and the centroid of X-ray emission. The filled histogram represents the distribution of offsets obtained in this work adding the forced identifications. The step-like histograms in red, green, and black represent the offset distributions for structures with cross-matches based on identifications from this work, CAMIRA, and WH21, respectively. The blue line indicates the two-component Gaussian fit derived using the data from this work (excluding forced identifications).
}
    \label{fig: offset}
\end{figure}

Figure \ref{fig: offset} shows the offset distributions for detections matched with the X-ray catalogs across the three algorithms. Following the methodology outlined in \citet{oguri18}, these distributions were fitted using the two-component Gaussian model described by \citet{oguri11} as

\begin{align}
f(d_{\rm{Off}}) &= f_{\mathrm{cen}} \cdot \frac{d_{\mathrm{off}}}{\sigma_1^2} 
\cdot \exp\left(-\frac{d_{\mathrm{off}}^2}{2\sigma_1^2}\right) \nonumber\\
&\quad + (1 - f_{\mathrm{cen}}) \cdot \frac{d_{\mathrm{off}}}{\sigma_2^2} \cdot 
\exp\left(-\frac{d_{\mathrm{off}}^2}{2\sigma_2^2}\right),
\end{align}

\noindent
where $d_{\rm{Off}}$ represents the transverse angular diameter distance between the BCG position identified by the cluster finder algorithm and the X-ray emission peak, $f_{\rm{cen}}$ denotes the fraction of well-centered clusters, $\sigma_{1}$ corresponds to the standard deviation of the well-centered population, and $\sigma_{2}$ is the standard deviation of the miscentered population. In Figure \ref{fig: offset}, we present the fitted model for our identifications. Table \ref{tab: offs} summarizes the parameters obtained for each distribution. The results show that the clusters identified  in this work are consistent within 1$\sigma$ with both CAMIRA and WH21, indicating that, regardless of the methodology, the overall accuracy in locating the BCG relative to the X-ray peak is comparable among the three algorithms. However, there is a more significant difference between CAMIRA and WH21, with the former showing a higher fraction of well-centered BCGs.

\begin{table}
\centering
\caption{Two-component Gaussian model fitted parameters for the distribution of the offset between the position of the BCG and X-ray centroid.}
\scriptsize
\begin{tabular}{cccc}
\hline
  & $f_{cen}$       & $\sigma_{1}$    & $\sigma_{2}$    \\
\hline
This work (w/ forced) & 0.39 $\pm$ 0.10 & 0.032 $\pm$ 0.003 & 0.31 $\pm$ 0.04 \\
This work             & 0.46 $\pm$ 0.06 & 0.033 $\pm$ 0.002 & 0.32 $\pm$ 0.03 \\
CAMIRA                & 0.50 $\pm$ 0.04 & 0.032 $\pm$ 0.001 & 0.24 $\pm$ 0.01 \\
WH21                  & 0.41 $\pm$ 0.04 & 0.034 $\pm$ 0.002 & 0.38 $\pm$ 0.02 \\
\hline
\end{tabular}
\label{tab: offs}
\end{table}

%-------------------------------------------------------------------
\section{Conclusions} \label{sec: conclusions}

We present a new catalog containing 16,007 candidates of (proto)clusters of galaxies selected from the photometric footprint of the HSC-SSP Wide survey, covering the redshift range $\rm 0.1 < z_{phot} < 2$. To extend the search to this high-redshift limit, we implemented new photometric redshift estimates using deep learning techniques in combination with a novel training sample constructed from spectroscopic data available in the HSC-SSP Wide PDR3 database. This includes objects with high-accuracy photometric redshifts from the COSMOS2020 catalog and incorporating mid-infrared measurements for galaxies with counterparts in the unWISE catalog. The dispersion of these photometric redshift estimates is comparable to those obtained by \citet{wh21}, which rely solely on objects matched with unWISE data.

Our selection process follows the methods described in detail by V25a, beginning with the identification of the dominant galaxy of the structure, i.e., (proto)BCGs. These are selected as massive galaxies located in regions of high galaxy density. Testing this method on mock catalogs designed to emulate the HSC-SSP Wide observations shows that it can produce a sample of (proto)clusters with approximately 90\% purity with $\sim$ 65\% (proto)BCGs being correctly assigned. The primary source of contamination arises from massive galaxies residing in halos with masses around $\rm \sim 10^{13.5} \ M_{\odot}$, i.e., galaxy groups. Once the positions of the central galaxies are identified, additional member galaxies are selected based on their proximity to the (proto)BCG and the expected stellar mass for galaxies at the photometric redshift of the structure.

Our findings indicate broad agreement between different catalogs generated by similar algorithms, but also reveal significant discrepancies, with up to 40\% of unique candidates being identified even at lower redshifts $\rm 0.1 < z_{phot} < 0.45$. These differences highlight that cluster candidate selection using diverse techniques leads to distinct discoveries, emphasizing that such catalogs are complementary rather than redundant.

Our analysis shows that the high-redshift candidates ($\rm z_{phot} > 1.5$) are located at prominent peaks in the stellar mass density field. Moreover, these systems host galaxy populations that are, on average, brighter and redder than field galaxies at the same redshift slice. This indicates that these candidates correspond to structures that have already undergone a certain degree of evolutionary processing.

With the start of operations of the Prime Focus Spectrograph instrument, the catalog of (proto)cluster of galaxies candidates produced in this work provides an additional valuable resource for selecting potential targets for spectroscopic follow-up, which will help to improve our comprehension of galaxy evolution in dense environments since $\rm z_{phot} \sim 2$.

%\texttt{Astropy} \citep{astropy, astropy2018}, 
%\software{\texttt{corner} \citep{corner2016}, \texttt{gala} \citep{gala2017}, \texttt{jupyter} \citep{jupyter2016}, \texttt{matplotlib} \citep{matplotlib}, \texttt{NumPy} \citep{numpy}, \texttt{pandas} \citep{pandasSoftware}, \texttt{SciPy} \citep{scipy}, \texttt{scikit-learn} \citep{scikit-learn}, \texttt{TOPCAT} \citep{TOPCAT2005}.
%}

\section*{Acknowledgements}
MCV acknowledges the Fundaç\~ao de Amparo à Pesquisa do Estado de S\~ao Paulo (FAPESP; 2021/06590-0) for supporting his PhD and Research Internship Abroad at the Department of Astrophysical Sciences, Princeton University. He also thanks the Department of Astrophysical Sciences at Princeton University for its financial support in making this internship possible. LSJ acknowledges the support from CNPq (308994/2021-3) and FAPESP (2011/51680-6). E. V. R. L acknowledges the support from CAPES (88887.470064/2019-00) and FAPESP (2024/15229-8). PA-A thanks the Coordenaç\~ao de Aperfeiçoamento de Pessoal de Nível Superior – Brasil (CAPES), for supporting his PhD scholarship (project 88882.332909/2020-01).

This paper is based on data collected at the Subaru Telescope and retrieved from the HSC data archive system, which is operated by Subaru Telescope and Astronomy Data Center at National Astronomical Observatory of Japan. The Hyper Suprime-Cam (HSC) collaboration includes the astronomical communities of Japan and Taiwan, and Princeton University. The HSC instrumentation and software were developed by the National Astronomical Observatory of Japan (NAOJ), the Kavli Institute for the Physics and Mathematics of the Universe (Kavli IPMU), the University of Tokyo, the High Energy Accelerator Research Organization (KEK), the Academia Sinica Institute for Astronomy and Astrophysics in Taiwan (ASIAA), and Princeton University. Funding was contributed by the FIRST program from Japanese Cabinet Office, the Ministry of Education, Culture, Sports, Science and Technology (MEXT), the Japan Society for the Promotion of Science (JSPS), Japan Science and Technology Agency (JST), the Toray Science Foundation, NAOJ, Kavli IPMU, KEK, ASIAA, and Princeton University. 

This paper makes use of software developed for the Rubin Observatory Large Synoptic Survey Telescope. We thank the LSST Project for making their code available as free software at  http://dm.lsst.org

The Pan-STARRS1 Surveys (PS1) have been made possible through contributions of the Institute for Astronomy, the University of Hawaii, the Pan-STARRS Project Office, the Max-Planck Society and its participating institutes, the Max Planck Institute for Astronomy, Heidelberg and the Max Planck Institute for Extraterrestrial Physics, Garching, The Johns Hopkins University, Durham University, the University of Edinburgh, Queen’s University Belfast, the Harvard-Smithsonian Center for Astrophysics, the Las Cumbres Observatory Global Telescope Network Incorporated, the National Central University of Taiwan, the Space Telescope Science Institute, the National Aeronautics and Space Administration under Grant No. NNX08AR22G issued through the Planetary Science Division of the NASA Science Mission Directorate, the National Science Foundation under Grant No. AST-1238877, the University of Maryland, and Eotvos Lorand University (ELTE) and the Los Alamos National Laboratory.

\vspace{5mm}

\software{Numpy \citep{harris2020}, Pandas \citep{reback2020}, Scipy \citep{scipy}, Matplotlib \citep{hunter2007}, Astropy \citep{astropy}, Tensorflow \citep{tensorflow2015}, Keras \citep{chollet18}, Sklearn \citep{pedregosa2011}}

\appendix
\counterwithin{figure}{section}
\counterwithin{table}{section}

\section{High-z candidates subset}
\label{apx: highz_sample}

\begin{table}[ht]
\caption{Example of high-redshift ($z > 1.5$) cluster candidates identified in this work. Superscripts refer to the column descriptions provided below the table.}
\centering
\begin{tabular}{lllllllll}
\toprule
$RA^{a} \ [deg]$ & $Dec^{b} \ [deg]$ & $i_{mag}^{c}$ & $\delta \rho^{d}$ & $P_{dominant}^{e}$ & $\lambda^{f}$ & $z_{cl}^{g}$ & $\log(M_{halo}/M_{\odot})^{h}$ \\
\midrule
179.54844 & 3.99159 & 23.11 & 4.02 & 0.87 & 18.0 & 1.52 & 14.0 $\pm$ 0.1 \\
244.80646 & 43.06906 & 23.65 & 3.65 & 0.82 & 12.0 & 1.63 & 13.7 $\pm$ 0.1 \\
129.57409 & 2.10479 & 24.09 & 2.47 & 0.56 & 10.0 & 1.74 & 13.6 $\pm$ 0.1 \\
212.81541 & 43.56201 & 23.73 & 1.96 & 0.43 & 12.0 & 1.87 & 13.7 $\pm$ 0.1 \\
188.55843 & 2.71758 & 23.42 & 3.62 & 0.81 & 11.0 & 1.56 & 13.6 $\pm$ 0.1 \\
334.63814 & -0.71778 & 23.66 & 2.44 & 0.56 & 13.0 & 1.51 & 13.7 $\pm$ 0.1 \\
2.86400 & 1.50079 & 23.41 & 2.56 & 0.59 & 13.0 & 1.51 & 13.7 $\pm$ 0.1 \\
159.27691 & 3.52094 & 23.81 & 1.68 & 0.37 & 12.0 & 1.86 & 13.7 $\pm$ 0.1 \\
353.55651 & 2.64799 & 22.91 & 1.93 & 0.43 & 11.0 & 1.50 & 13.6 $\pm$ 0.1 \\
147.53107 & 4.44377 & 21.90 & 1.89 & 0.42 & 18.0 & 1.53 & 14.0 $\pm$ 0.1 \\
178.73746 & 3.47801 & 23.78 & 1.70 & 0.37 & 12.0 & 1.52 & 13.7 $\pm$ 0.1 \\
30.59459 & 1.41142 & 23.83 & 1.84 & 0.40 & 11.0 & 1.67 & 13.6 $\pm$ 0.1 \\
146.12104 & 4.54147 & 23.59 & 2.27 & 0.51 & 14.0 & 1.54 & 13.8 $\pm$ 0.1 \\
223.19057 & 42.17150 & 22.31 & 3.03 & 0.70 & 11.0 & 1.51 & 13.6 $\pm$ 0.1 \\
162.80712 & -0.34168 & 23.71 & 3.96 & 0.86 & 11.0 & 1.52 & 13.6 $\pm$ 0.1 \\
148.72475 & 4.67181 & 23.47 & 1.45 & 0.31 & 12.0 & 1.89 & 13.7 $\pm$ 0.1 \\
212.53725 & -1.71297 & 23.98 & 1.97 & 0.44 & 12.0 & 1.63 & 13.7 $\pm$ 0.1 \\
147.79017 & 1.73523 & 24.02 & 3.00 & 0.69 & 12.0 & 1.51 & 13.7 $\pm$ 0.1 \\
224.71253 & 42.10175 & 22.55 & 1.67 & 0.36 & 12.0 & 1.72 & 13.7 $\pm$ 0.1 \\
144.09236 & 4.63590 & 23.53 & 1.77 & 0.39 & 11.0 & 1.54 & 13.6 $\pm$ 0.1 \\
244.34458 & 42.91256 & 24.15 & 2.08 & 0.47 & 13.0 & 1.54 & 13.7 $\pm$ 0.1 \\
192.35331 & 2.72834 & 23.09 & 3.27 & 0.75 & 12.0 & 1.51 & 13.7 $\pm$ 0.1 \\
353.88123 & 2.49429 & 23.14 & 4.80 & 0.94 & 13.0 & 1.53 & 13.7 $\pm$ 0.1 \\
224.52003 & 42.92812 & 24.00 & 1.72 & 0.38 & 11.0 & 1.60 & 13.6 $\pm$ 0.1 \\
21.34049 & -0.35689 & 22.95 & 2.63 & 0.60 & 13.0 & 1.58 & 13.7 $\pm$ 0.1 \\
\bottomrule
\end{tabular}
\label{tab: appendix}
\tablecomments{$^{a}$ Right Ascention of the dominant galaxy candidate, i.e., (proto)BCG; $^{b}$ Declination of the dominant galaxy candidate; $^{c}$ i-band magnitude of the dominant galaxy candidate; $^{d}$ stellar mass contrast density associated with the dominant galaxy candidate as defined in V25a; $^{e}$ Probability associated to the dominant galaxy candidate being correctly assigned; $^{f}$ Richness of the (proto)cluster candidate as defined in V25a; $^{g}$ Redshift of the (proto)cluster candidate; $^{h}$ (Proto)cluster halo mass estimate as defined in V25a.} 
\end{table}

%\begin{figure*}[h!]

%	\includegraphics[width=\textwidth]{figures/UTS_COSMOS2020_dist1.png}
%    \caption{  
%}
%    \label{fig: uts_cosmos_1}
%\end{figure*}

%\begin{figure*}[h!]

%	\includegraphics[width=\textwidth]{figures/UTS_COSMOS2020_dist2.png}
%    \caption{  
%}
%    \label{fig: uts_cosmos_2}
%\end{figure*}

%\clearpage
%\bibliography{bibliography.bib}{}
%\bibliographystyle{aasjournal}

\end{document}